/

# Copper red glazes: a coating with two families of particles

**P.A. Cuvelier, C. Andraud, D. Chaudanson, J. Lafait, S. Nitsche**

**Abstract** In order to explain the origin of the deep red color of copper glazes on ceramics, a ceramist has elaborated, by firing under reducing atmosphere, a significant number of tiles. The analysis of the structure and composition of a representative sample by TEM and EELS, followed by an optical characterization and an optical modeling using the radiative transfer approach (four-flux model) have enabled a comprehensive explanation of the origin of the color in these glazes. The presence of two families of copper nanoparticles in the vitreous matrix. The first, purely absorbing, of diameter 10-50 nm, essentially creates color by a substractive process. The second, due to its larger diameter, 100 nm or more, mainly acts on color by scattering of the visible light. Both act competitively in the layer. A color chart of all the hues which can be reached by this technique has eventually been theoretically calculated.

## 1. Introduction

Copper red in glazed ceramics is a wonderful color searched out by generations of ceramists, often obtained, never well controlled, never seriously analyzed under a scientific point of view.

P.A. Cuvelier corresponding author ,
R2C-rougedecuivre.com, 37, impasse de la Vielle,
13009 Marseille - France
e-mail : cuvelier.pa@wanadoo.fr

C.Andraud, J. Lafait
Institut des Nanosciences de Paris – UMR 7588 – CNRS – Université Pierre et Marie Curie - Campus Jussieu - Case 840 – 4, place Jussieu – 75252 Paris cedex 05 – France

D. Chaudanson, S.Nitsche
Aix Marseille Université, CINaM, UPR CNRS 3118, Campus Luminy, F 13288 Marseille cedex 09, France

It began to be widely used as splashes on plain glaze or as underglaze decoration, approximately at the 12th century in China [1], followed by major success in this country as a monochrome color at the 15th and 17th centuries, before its diffusion in Europe and all around the world since the 19th century. It shares part of its optical properties with other materials of the cultural heritage, gold ruby glass and luster ceramics (like the well known Gubbio production at the 16th century [2] due to the common presence in a glass matrix of metallic nano-particles which exhibit a specific color related to surface plasmon excitation in the metallic grains. The noble metal can be copper, gold or silver. Nevertheless, in ruby glass, the material remains transparent, like stained glass, while in luster ceramics it is the interference effect in a multilayer structure which [3,4], as a first approximation, achieves the iridescent color. In both cases, contrarily to the case of copper red glazes, light scattering never occurs in the bulk of the matrix or the coating. These three examples have in common the fact that the color is produced by a structural effect. It is, in fact, one aim of this paper to establish this explanation for copper red glazes, on solid physical basis. Although it is generally admitted that metallic copper is at the origin of the color, different controversial explanations [5-9] have indeed up to now been evoked, sometimes without any serious support, either the presence of copper nanoparticles, or the presence of cuprous oxide ($Cu_2O$). At our knowledge, the relation between the color of the glaze and the physico-chemical state of the particles and moreover, the intimate structure of the layer, has never been treated. We wanted to go much further and try to answer precisely (qualitatively and quantitatively) simple (!) questions like: (i) what can make the color lighter or darker (or change the saturation)?; (ii) what can change the hue?; (iii) what can enhance the brightness? ;etc…

These questions are also the first step of a comprehensive work to help the ceramist or the potter, to better control his technique of elaboration of this very specific glaze which remains delicate to achieve because still poorly known. With this aim, one of us (P.A.C.), engineer and ceramist, has elaborated a substantial number (several thousands) of copper red glaze tiles by using the same technique. One of these tiles has been selected for its visual aspect: homogeneous and representative of the average hue. Its structure and composition have been studied in the depth of the glaze by TEM (transmission electron microscopy) and EELS (electron energy loss spectroscopy). Due to the scattering of light, both the specular and diffuse reflectance and transmittance of the glaze have been measured in the visible spectrum. This technical part of the experiments is presented in section 2. The results of the TEM and EELS experiments are given and discussed in section 3. These results, combined with the fact that all the tiles scatter the visible light, enabled us to make the choice of a theory able to account for the optical properties of this kind of sample, the radiative transfer approach, presented in section 4. In this section, we also give the results of theoretical optical simulations on layers containing Cu or Cu oxide particles, in order to check whether optical measurements alone were able to discriminate between these two controversial hypothesis concerning the composition of the glaze. We also present optical simulations as a function of the size of the inclusions, in relation with the results of the TEM experiments. At the end of this section, one has already a precise idea of the possible links between the morphology of a glaze layer and its color. We have then all the tools for comparing the optical measurements on the tile to the predictions of the model (section 5). Considering that our model has been validated and is therefore able to predict the optical properties and the color of Cu red glaze tiles presenting some variation in their structure and composition, we have calculated the colors which could be obtained within realistic variations of these parameters. This resulted a theoretical color chart presented in section 6. Section 7 is devoted to a discussion of the ensemble of these results and to some points which merit to be followed on, while the conclusion (section 8) brings together and summarizes the main decisive results of this study.

## 2. Experimental techniques

2.1. Technique of elaboration of the Cu red glaze tile and preparation of the samples

The general technique of elaboration of these Cu red glazes consists in coating a porcelain tile (first fired at around 900°C) with a sauce composed of a mixture of feldspar, chalk, talc, kaolin and silica and, sometimes, baryum oxide and /or boron fritt, to get a glaze. It is added to this mixture a low percentage of copper carbonate and SnO which are key components of the recipe. The coated porcelain is then fired at 1280°C in a potter gas oven under reducing conditions during the main part of the firing: typically 2% CO atmosphere for 2 hours from 950°C to 1180°C, then 1 hour at 0.5% CO till 1280°C, then neutral for half an hour, before gas shutdown and natural cooling (see example of test tiles in Fig. 1a).

The tile that we have selected for this study has been prepared with the following composition of the glaze mixture (values in mass percentage): $Na_2O$=1.7%, $K_2O$=7.7%, MgO=1.5%, $Al_2O_3$=13.0%, $SiO_2$=76.2%, with the addition of $CuCO_3$=0.6% and SnO=0.6%. The firing cycle was close to the one mentioned above, and the color of the tile after firing is a deep red (Fig. 1b).

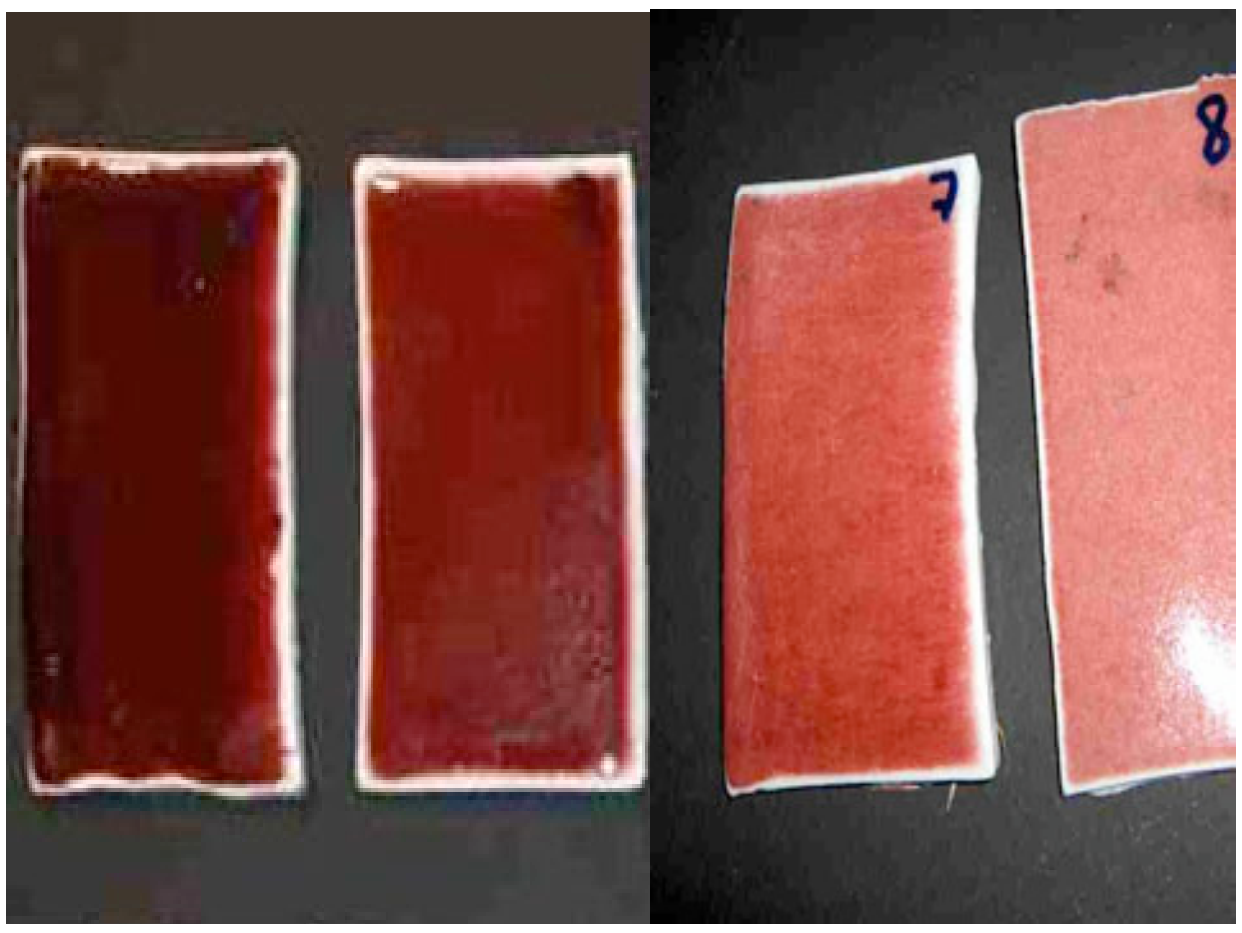

**(a)**

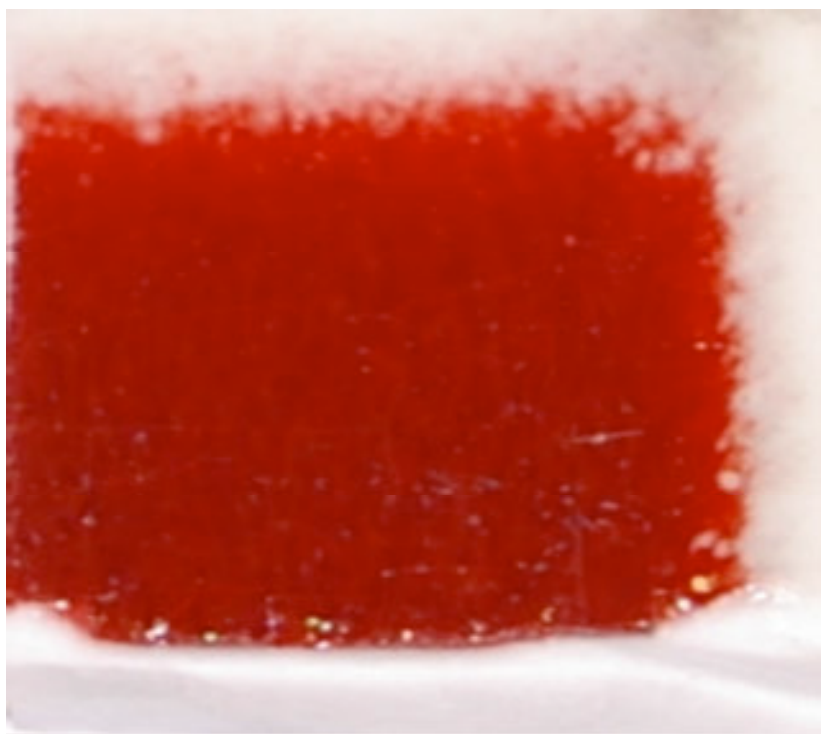

**(b)**

**Fig. 1 a-** *Examples of test tiles (8 cm x 4 cm)*
**b-** *Enlargement of a portion (15 mm x 16 mm) of the tile B23 studied in this paper*

The cross section of the tile B23 (Fig. 2a ) shows a rather uniform color in the middle part of the glaze, between two colorless zones, which is usual for copper red glazes.
The picture (Fig. 2b) of the layer 1 (thickness 35 µm) in the middle of the colored part of the glaze confirms the uniformity, and shows bubbles which are also usual in this kind of glaze.

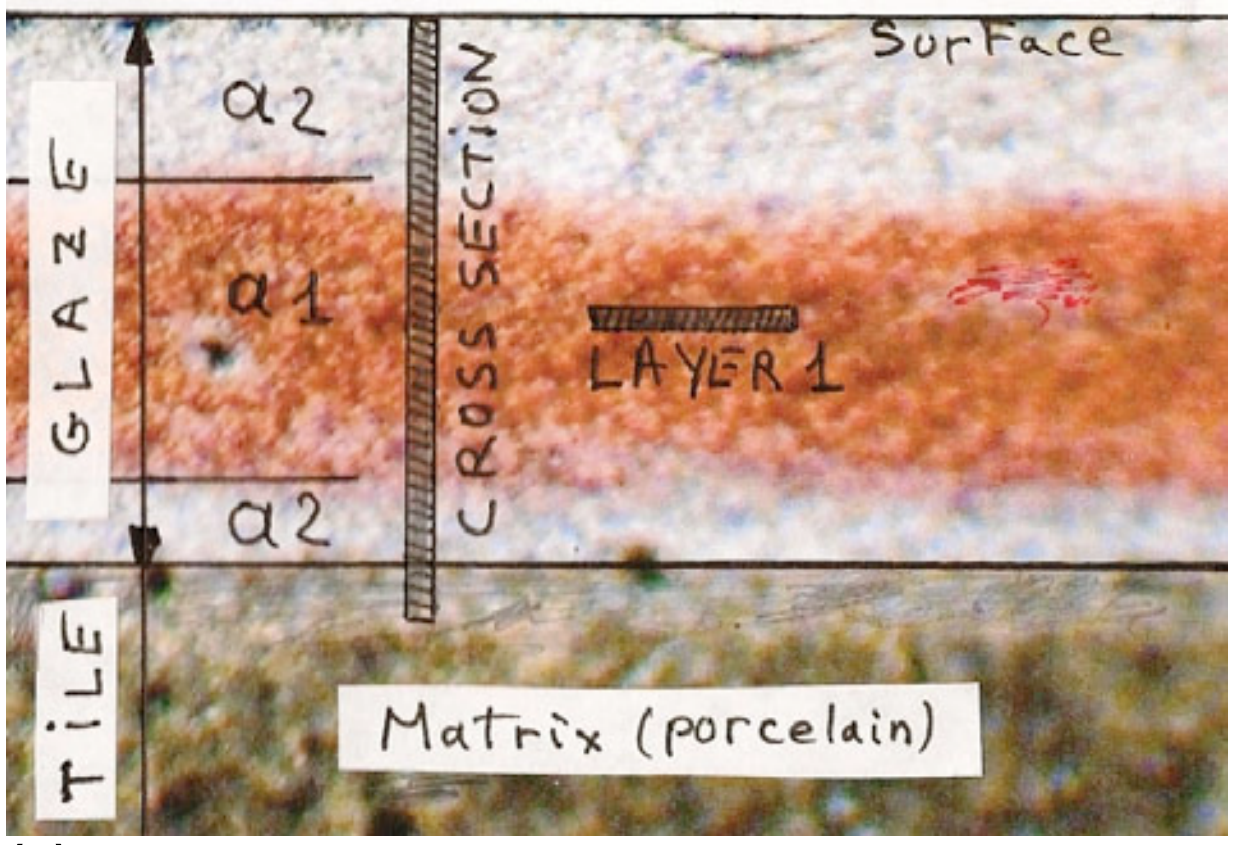

**(a)**

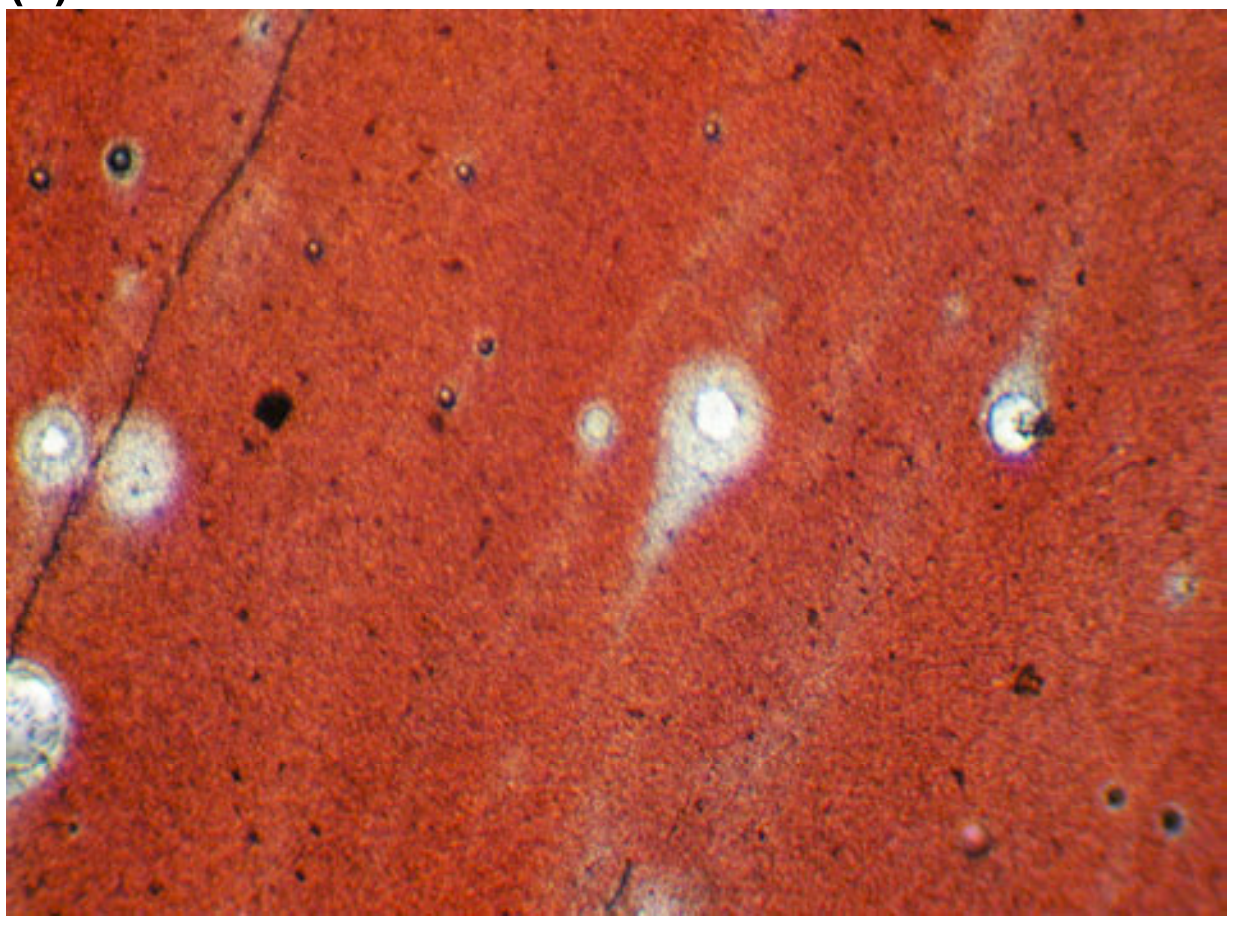

**(b)**

**Fig. 2**
**a-** *Picture of a cross section of the glazed tile B23*
*- GLAZE: thickness 0.3 mm*
*- a1 - colored area (150 µm), a2 - colorless areas*
*- TILE: made of porcelain (thickness 4 mm )*
*- CROSS SECTION: used for TEM and EELS*
*- LAYER 1: layer with thickness 35 µm used for optical, TEM and EELS analysis*
**b-** *Picture of layer 1 (6 mm x 4 mm)*

In order to perform optical analysis of the glaze, we have cut four layers parallel to the porcelain body that have been polished for optical measurement (Fig. 3a):
- layer 1 already shown Fig. 2 (colored thickness 35 µm) is located in the middle of the red zone, and was also used for TEM examination after optical measurements
- layer 2, is representative of the whole glaze (149 µm)
- layer 3 (114 µm) represents the part near the surface, and layer 4 (72 µm) near the porcelain.

2.2. TEM and EELS Measurements

Fig. 2a shows the location of the two thin sections which have been prepared, then ion milled with a Precision Ion Polishing System (PIPS) from GATAN, for TEM and EELS observations:
-The first section, located in the "LAYER 1"(in the red zone), is parallel to the bottom of the glaze. Its surface usable for observation is approximately 1 mm x 0,5 mm.
-The second one, located in the "CROSS SECTION", is perpendicular to the bottom of the glaze and permits the examination from the surface of the glaze to the interface glaze /porcelain.
TEM measurements were performed on both sections:
- at first, with a JEOL 2000FX electron microscope, as well as chemical microanalysis (X-ray spectroscopy )
-second, with a high resolution TEM JEOL 3010,
Then, EELS studies have been carried out on a TEM JEOL 2010F, equipped with an EELS GATAN system. The TEM size of the electron beam was adjusted at less than 4 nm in diameter to have a very good spatial resolution.

2.3. Optical measurements

2.3.1. Reflectance measurements

Two kinds of measurements have been performed:
On the tiles: specular measurements under normal incidence have been made with a spectrometer Minolta 3200, the sample being illuminated under roughly lambertian conditions by a white source. Following the second law of thermodynamics, the reflectance thus measured is equal to the diffuse reflectance of the sample illuminated under normal incidence ($R_{cd}$).
On the glaze layers put on a black, non-reflecting, background: specular reflectance measurements were performed using a spectrometer Ocean Optics 4000, (equipped with a halogen lamp) collimated illumination at 10°, measurement at 30°. Fig. 3b presents the reflectance spectra of the tile and of 4 layers in the glaze.

2.3.2 Transmission measurements

Normal specular transmittance ($T_{cc}$) measurements were performed on the four above layers, with the same Ocean Optics 4000 equipment (see Fig. 3c)

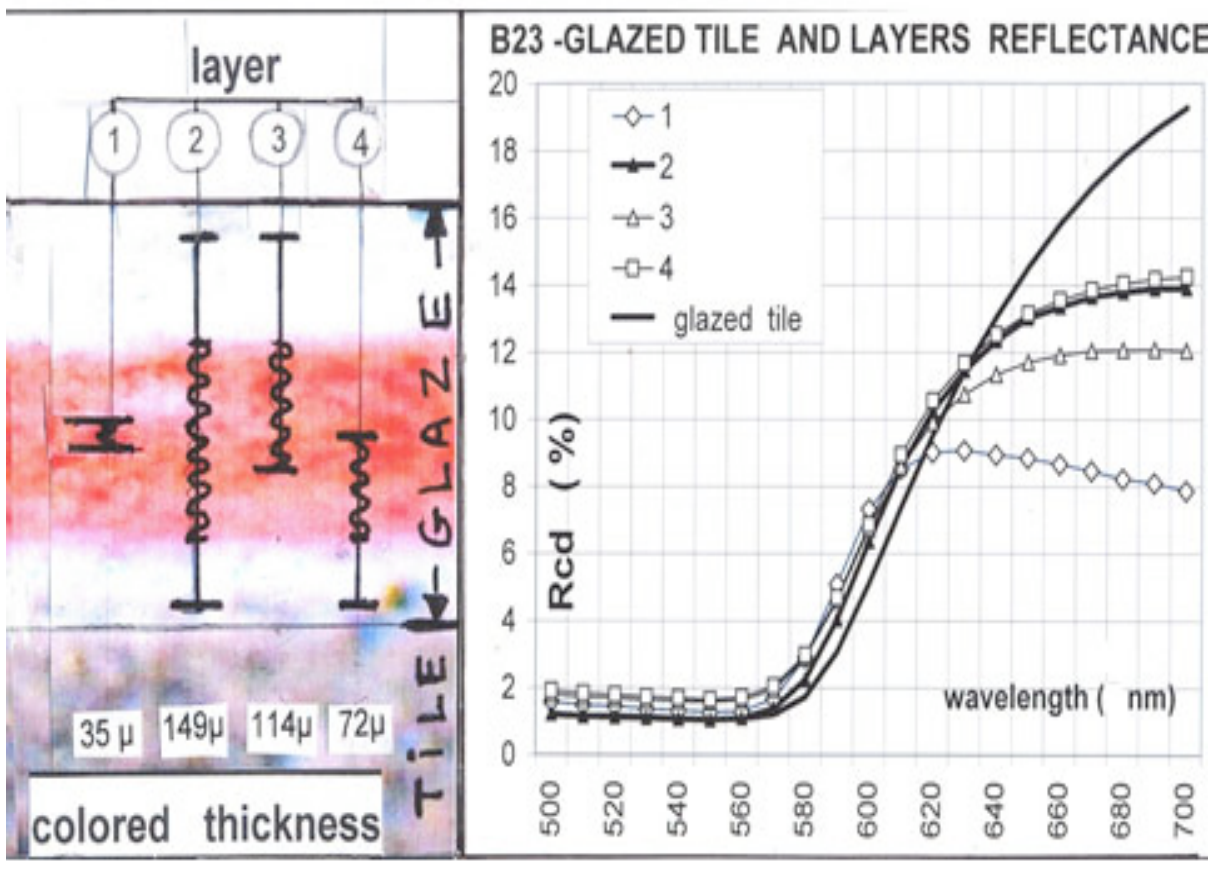

(a) (b)

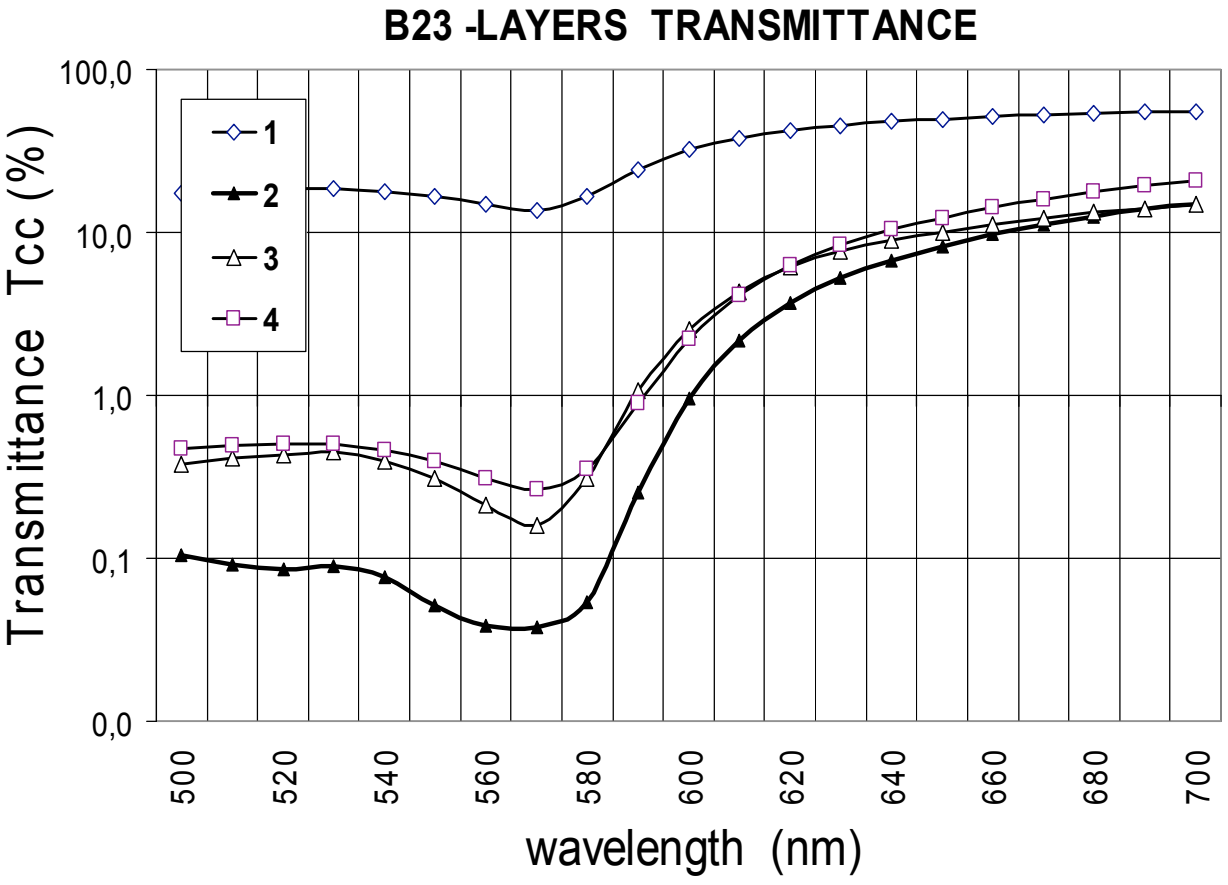

(c)

**Fig. 3** *Optical measurements on tile B23 and glaze layers*

**a** - *Sample area, and colored thickness, of the four layers prepared from the tile B23*
**b** - *Measured reflectance $R_{cd}$ of tile and of glaze layers on black background*
**c** - *Measured specular transmittance $T_{cc}$ of the layers*

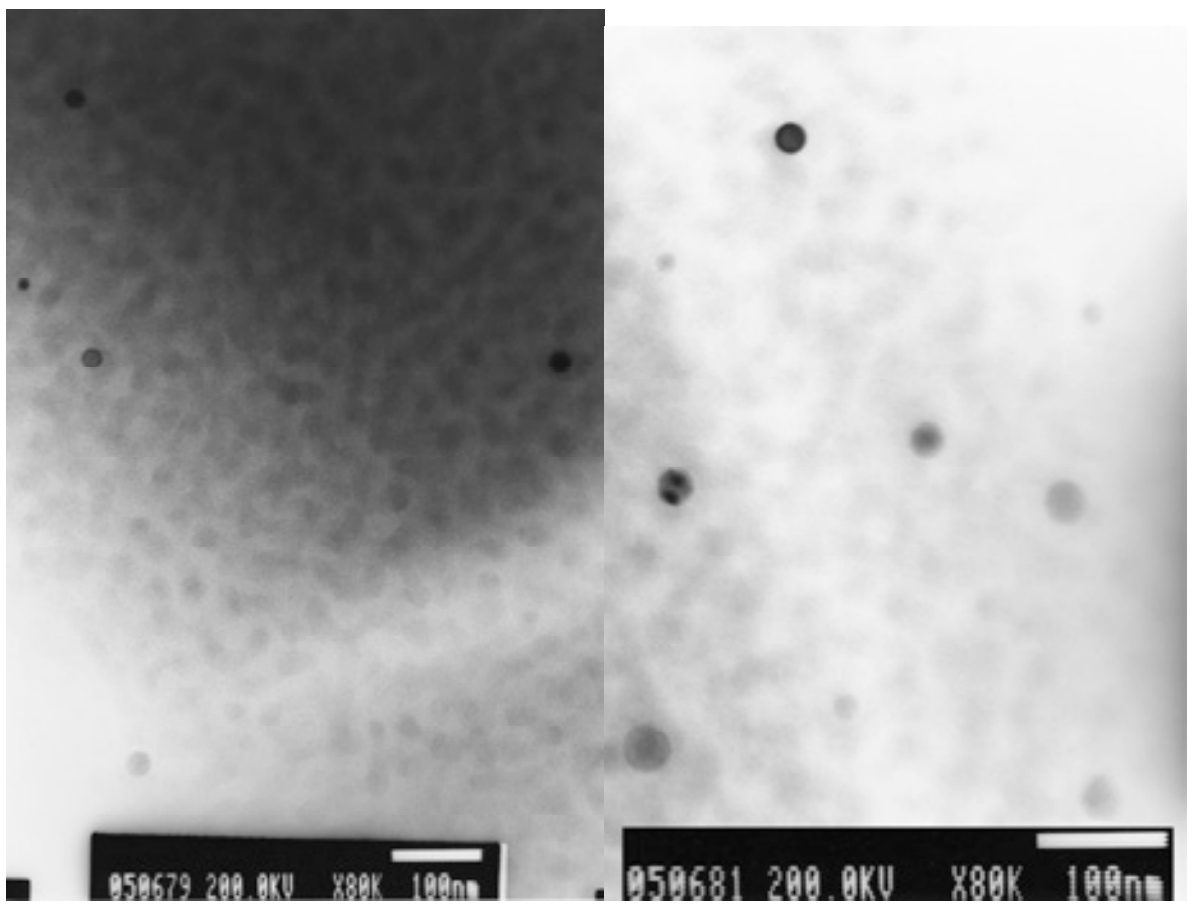

(a)

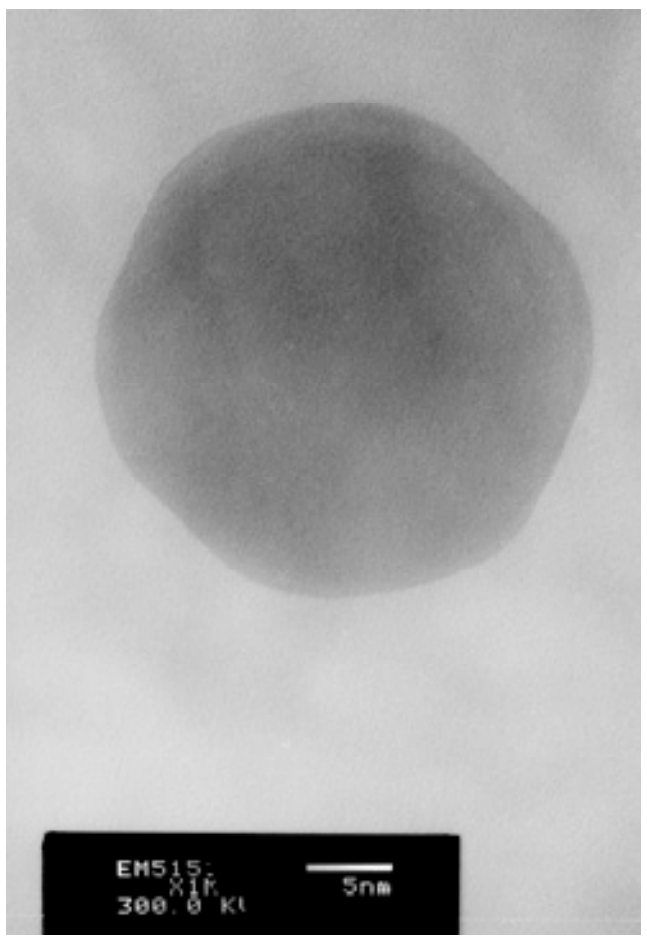

(b)
**Fig. 4** *TEM Observation of particles in layer 1*
**a** - *Pictures showing only the smallest ones*
**b** - *Example of a faceted small particle*

### 3-TEM and EELS Studies

3.1. Analytical TEM JEOL 2000FX studies

3.1.1. Result of observations

A global observation of the two thin layers (parallel and normal to the bottom of the glaze) shows the existence of two particles families with different sizes.
First family (Fig. 4):

- Size: there is no particle with a diameter less 10 nm, and the average size is around 30 nm.
- Shape and nature: all the particles are rather spherical, regular for those of less than 30-40 nm, and frequently faceted (Fig. 4b) for the biggest (crystalline habit).
- Number and distribution: In the thin section parallel to the bottom of glaze, the particles seem relatively homogeneously distributed and numerous enough to allow, from a few pictures, an evaluation of their distribution and also, roughly, of their total volume.

Figure 4c shows the histogram made from 6 pictures taken in this section (100 nm thick) which leads to a mean diameter of around 25 nm, (and a diameter of 30 nm when calculated from the mean volume of particles).The particle volume fraction in the glaze can be thus estimated around 12 x $10^{-4}$, value to be compared to the ratio of introduced copper, 9 x $10^{-4}$.

It should be noted also that the observation of the perpendicular thin cross section reveals that areas close to the surface of the glaze are little rich in particles.

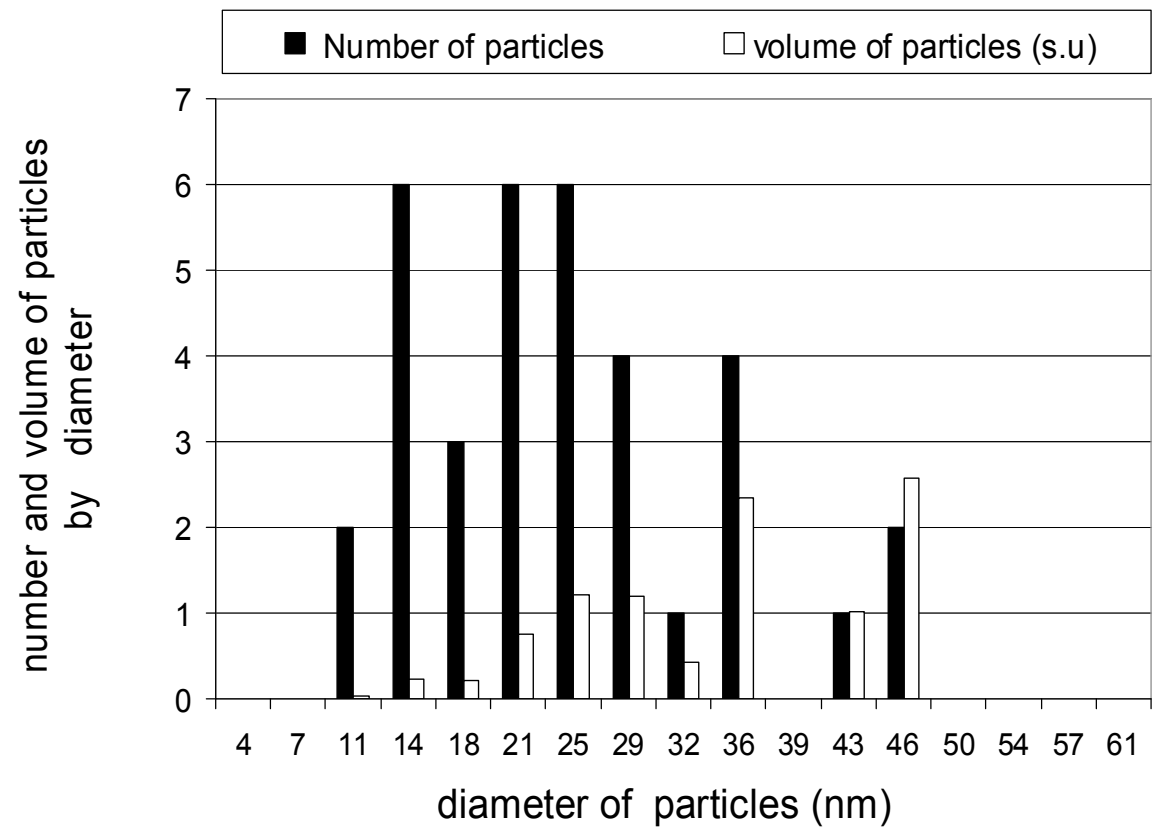

**Fig. 4c** *Histogram of small particles (diameter and volume) from a few photos in the layer 1 parallel to the bottom of glaze*

Second family (Fig. 5):

Size: diameter between 70 and 150 nm
Shape and nature: spherical shape particles frequently faceted (crystalline habit)
Number and distribution: these particles are in low number, and therefore difficult to quantify,

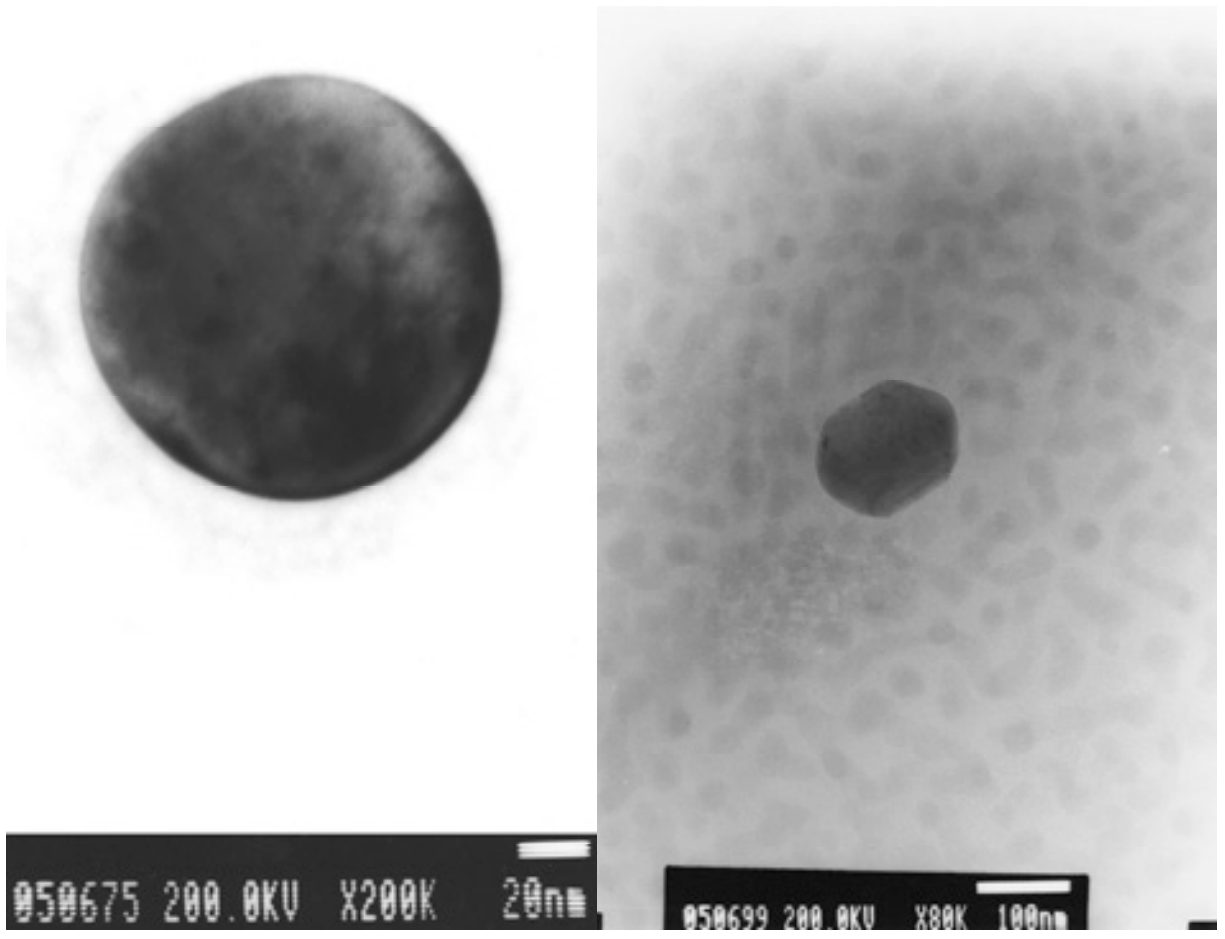

**Fig. 5** *Examples of large particles*

3.1.2. Result of chemical microanalysis of the glaze by EDX (energy dispersive X-ray spectroscopy)

a - Matrix: Measurements made in five points of the thin section parallel to the bottom of the glaze show (Table 1) an important spatial dispersion of composition coming probably from glaze inhomogeneities at the scale of the measure (each one corresponds to an analysis volume of 50nm side approximately). This effect lowers the precision of the mean value (see the case of Al, for instance), and also the coherence with values issued from glaze recipe.

More specifically, we didn't determine the quantity of tin (0.6 % of tin oxide in mass in the initial mixture), of Na and K (7.5 % mass in total). For the last two elements, this is explained by their volatilization (usually called "distillation") under the electron beam. The almost absence of tin is more surprising but is consistent with a uniform concentration of this element in the vitreous matrix, together with a detection limit of about 0.5%.

One can also notice the presence of about 0.7 % of iron, as an impurity in basic products, not quoted in the recipe. This order of magnitude is usual for the industrial products used in the present work. The same comment can be done for the measured values of Ti and P.

Note also that copper is not detected in the matrix. It is consistent with the low introduced quantity (0.3 % in mass), and the fact that copper is found mainly in nanoparticles (see next paragraph) and that the EDX detection threshold is of the order of 0.5 %.

| % in mass | point 2 | point 4 | point 6 | point 7 | point 10 | **mean value glaze** | **Value from recipe ( K, Na Exclu-ded)** |
|---|---|---|---|---|---|---|---|
| **Si** | 82,2 | 64,9 | 77,4 | 68,6 | 74,4 | **73,5** | **68,0** |
| **Al** | 5,8 | 10,8 | 10,1 | 12,2 | 11,2 | **10,0** | **13,2** |
| **K** | 1,48 | 3,12 | 0,45 | 1,42 | 0,05 | **1,3** | **-------** |
| **Ca** | 9,6 | 17,9 | 9,67 | 14,7 | 9,42 | **12,3** | **16,7** |
| **Fe** | 0,83 | 0,77 | 0,52 | 0,74 | 1,05 | **0,78** | **0** |
| **Sn** | 0,08 | 0,01 | 0,01 | 0,03 | 0,1 | **0,05** | **0,4** |
| **P** | 0,03 | 0 | 0 | 0,01 | 1,82 | **0,37** | **0** |
| **Ti** | 0,03 | 0,15 | 0 | 0,01 | 0,04 | **0,05** | **0** |
| **Na** | | 0,26 | 0 | 0,31 | 0,04 | **0,12** | **-------** |
| **Mg** | | 2,04 | 1,78 | 2 | 1,82 | **1,53** | **1,7** |
| | 100 | 100 | 100 | 100 | 100 | **100** | **100** |

**Table 1** *JEOL 2000FX TEM chemical microanalysis results*

b-Particles: We have determined by chemical microanalysis (EDX), the elementary composition of all the particles observed in the glaze sample (more than 20 measurements of small and large particles), the size of the beam having been reduced for the analysis of the smallest. We found Copper, but never Tin or Iron.

However, we cannot discriminate the presence or absence of copper linked to the oxygen, because the particles are embedded in a matrix already containing oxygen.

3.2. High resolution TEM JEOL 3010 studies

3.2.1. Small particles (10 to 50 nm in diameter)

They seem to be monocrystalline, from the observation of the lattices fringes and of the diffraction patterns (Fig. 6). Furthermore, the distances between atomic planes (see Fig 6a and b) are characteristic of metallic copper: 2.07 Å for planes (111) and 1.21 Å for planes (110)

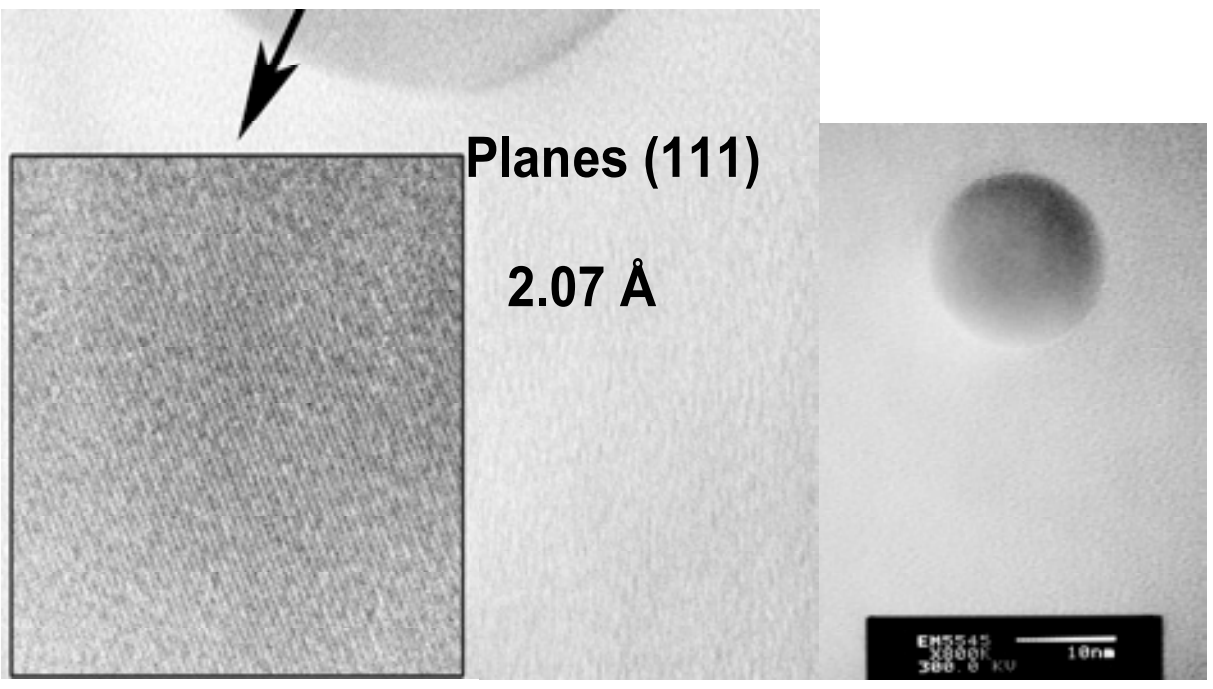

(a)

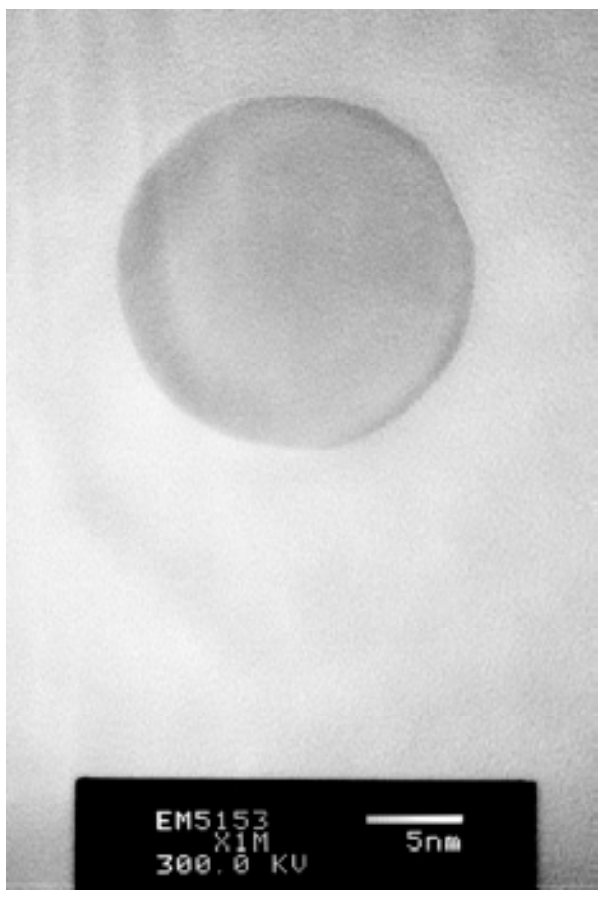

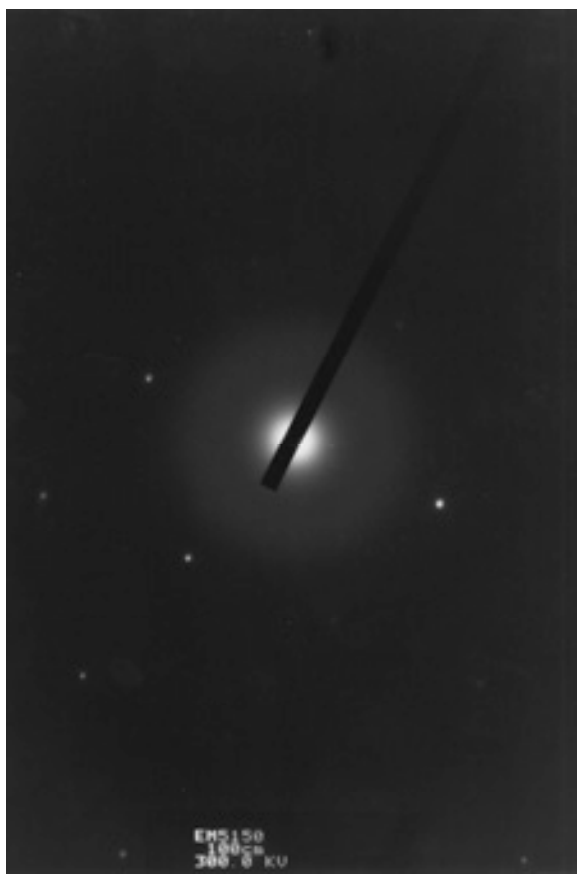

(b) (c)

**Fig. 6** *Small particles imaging, high resolution TEM.*
**a** - *Lattice fringes of Cu at 2.07 Å (111)*
**b** - *Lattice fringes of Cu at 1.21 Å (110)*
**c** - *Diffraction pattern of disoriented Cu particle*

3.2.2. Large particles

They seem (according to the imaging and diffraction- Fig. 7), to be also monocrystalline but with many defects, multi-twins, stacking fault, etc…These large particles are too thick (50 to 150 nm in diameter) to allow imaging of atomic planes.

For both sizes, the electron energy loss spectroscopy (EELS) should be able to confirm that small particles are metal copper and determine the composition of the large ones: oxide or metal.

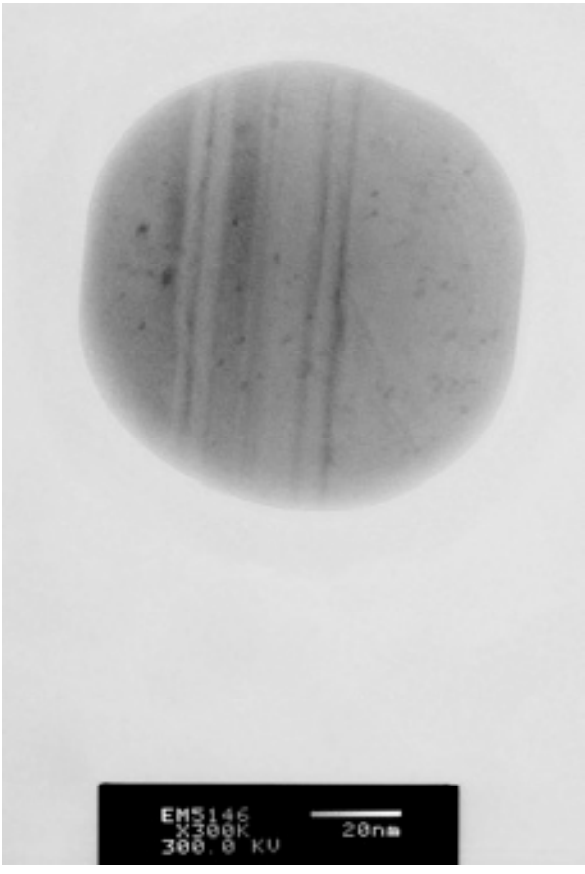

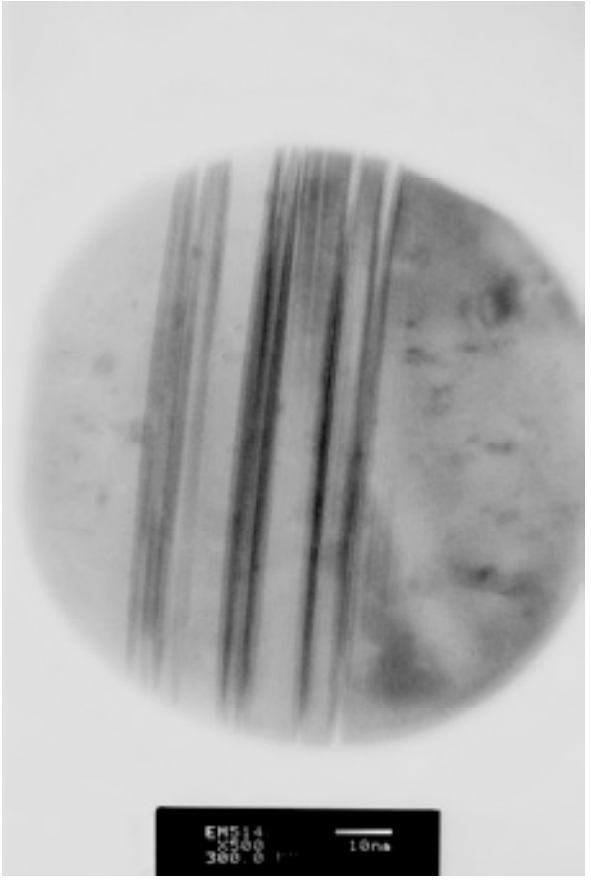

**(a)** **(b)**

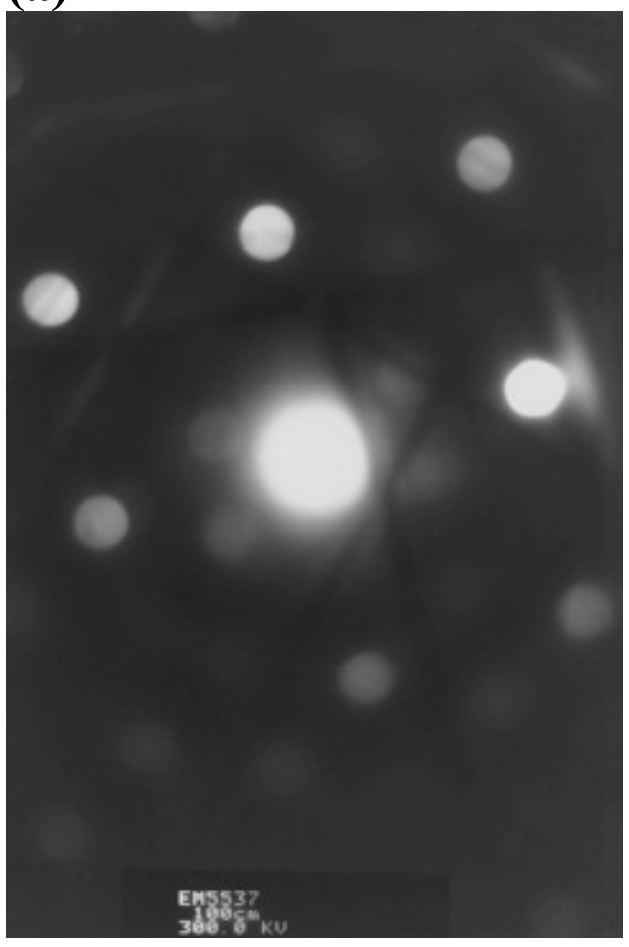

**(c )**

**Fig. 7** *Large particles imaging, high resolution TEM.*
**a, b** - *Examples of large particles*
**c -** *Diffraction pattern of disoriented Cu particle*

3.3. EELS studies

They have been carried out on a TEM ( JEOL 2010F), equipped with an EELS GATAN system. The TEM size of the electron beam was adjusted below 4 nm in diameter in order to get a very good spatial resolution.

3.3.1. On the thin section parallel to the bottom of the glaze:

- Small particles:
The analysis have been carried out on five particles of various diameters. Each time, the resulting spectrum was concluded to be pure metallic copper.
- Large particles:
The measurements have been carried out on two particles. Each time, the resulting spectrum corresponds to pure metallic copper. The scan from the center of the particle to the periphery has not shown the presence of copper oxide.

3.3.2. On the thin section perpendicular to the bottom of the glaze:

Many particles were analyzed by EELS (in their centre and their edge) by scanning all the thickness of the glaze layer. These measurements show only metallic copper, with the exception of one particle on which we have detected oxide (only at the periphery). This exception will not be considered as significant but confirms the ability of the method to detect $Cu_2O$. Iron was not detected, confirming thus the previous measurements

We found two families (in size) of particles, “small” and “large” and confirm the first JEOL 2000FX observations: particles are virtually absent close to the surface. This observation on cross section confirms the results obtained on the thin section parallel to the bottom of the glaze.

3.4. TEM concluding studies

One can conclude for the glaze of the B23 tile:
1-Particles are metallic copper, for all sizes and locations in the glaze.
2-Particles can be divided into two families:
- a small particles family, with a distribution in diameter of about 10 to 50 nm, centred at 30 nm, with no particle less than 10 nm. Their number is important and roughly corresponds to the mass of copper introduced in the initial sauce.
- a large particles family, with a distribution in diameter from 70 to 150 nm. They are difficult to quantify because their number is very low.
3-The distribution between the two families does not seem to depend on the depth in the glaze

These results allow an identification of the nature of the particles supposed to be at the origin of the color. However they only provide a rough estimate of the concentration of small particles, and they cannot estimate the concentration of large particles due to the scale of these observations.

As stated before we want to go beyond the simple knowledge of the physico-chemical state of particles by trying to link it to the color of the glaze by means of optical modeling.

## 4. Optical modeling: effect of copper and copper oxide particles

We here only envisage the effect of spherical particles, with the same composition and low enough concentration to consider the particles free from coherent optical interaction. It is the case of the glaze under consideration. The particles look roughly spherical and their volume ratio is of the order of $10^{-4}$.

4.1. Small non scattering particles

4.1.1. Homogeneous particles

When the diameter of the particles is less than around 50 nm, we will neglect the scattering phenomena and only consider the absorption which is mainly due to the surface plasmon excitation when the particles are metallic. We will use the absorption coefficient $\alpha(\lambda)$ of the glaze for describing the attenuation of the intensity $I(\lambda)$ of a monochromatic light propagating in a slab of the glaze of thickness $d$:

$$I(\lambda) = I_0(\lambda)\exp - \left[\alpha(\lambda) d\right] \quad (1)$$

$\alpha(\lambda)$ can be deduced experimentally from reflexion and transmission measurements on the glaze layer and on its substrate. Under the point of view of modeling, the absorption coefficient of such an heterogeneous medium composed of inclusions of size small compared to the wavelength, can be simply described at low concentration by the Maxwell-Garnett model, which leads to the following formula: (see for instance [10]):

$$\alpha(\lambda) = \rho_{vol} \cdot \frac{6\pi}{\lambda} \cdot \sqrt{\varepsilon_m} \cdot \mathrm{I}_m \left[ \frac{\tilde{\varepsilon} - \varepsilon_m}{\tilde{\varepsilon} + 2\varepsilon_m} \right] \quad (2)$$

Where
$\rho_{vol}$ = volume ratio of the inclusions (particles)
$\varepsilon$ = dielectric function of the matrix (glaze) assumed to be real
$\tilde{\varepsilon}$ = complex dielectric function of the inclusions
$\lambda$ = wavelength in vacuum or air (n=1)
We recall that the complex dielectric function $\tilde{\varepsilon}$ is related to the complex refractive index $\tilde{n}$ via: $\tilde{\varepsilon} = \tilde{n}^2$

When the particles are not too small (few tens of nanometers), the dielectric function of the metal can be described by the Drude model provided a classical size correction be applied to the electron damping coefficient $\gamma$ via

$$\gamma = \gamma_0 + v_f / r$$

where $v_f$ is the Fermi velocity of conduction electrons in bulk metal and $r$ the particle radius. $\gamma_0$ is the damping coefficient in the bulk, taken in this work equal to 0.15 eV for copper. At very small sizes, quantum corrections need to be introduced in the dielectric function of the metal. It is however not the case in our glazes.

Along this study, we used as dielectric function of Cu and $Cu_2O$ the currently admitted values given respectively by Palik [11] and Ito [12]. The index of the glaze has been taken equal to 1.5 which corresponds to

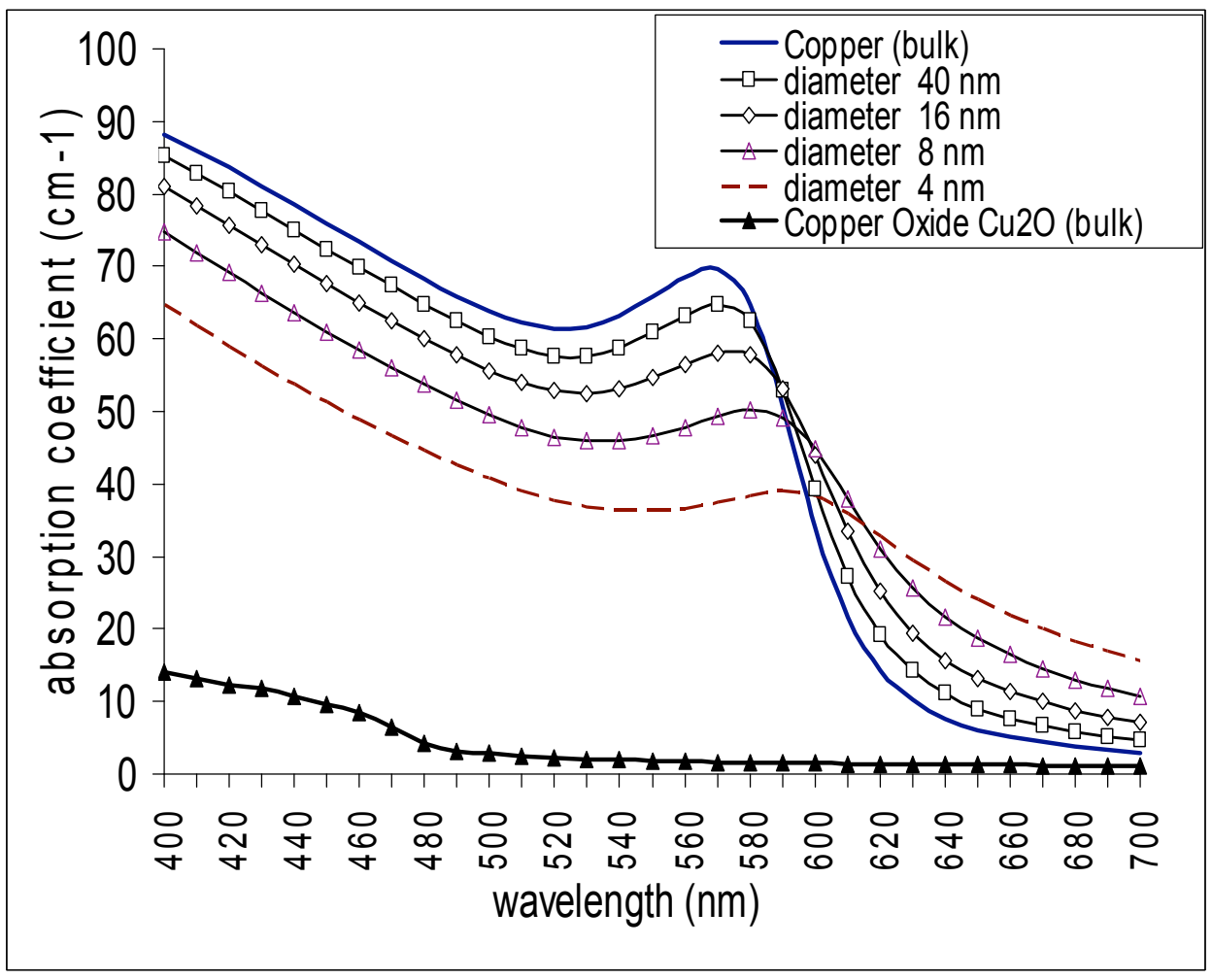

**(a)**

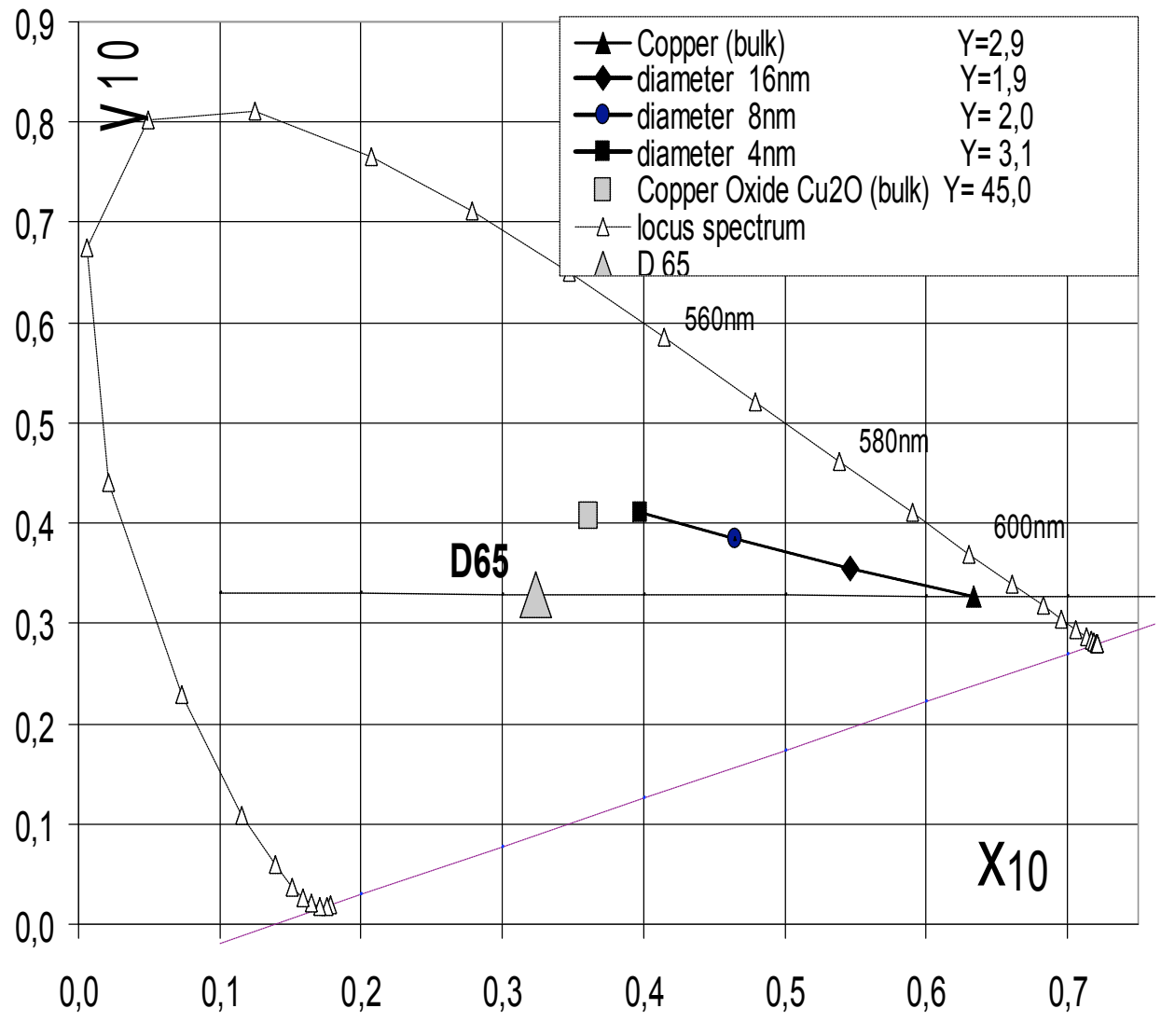

**(b)**

**Fig. 8** *Calculated absorption coefficient* ***(a)*** *and color coordinates x, y, Y* ***(b)*** *of a glaze containing Cu or $Cu_2O$ small particles ($\rho_{vol}$ =2 x$10^{-4}$, glaze thickness =150 μm)*

the average value measured on all the samples that we have elaborated and measured (ranging from 1.49 to 1.52).

By comparing the experimental absorption measured on our samples and the predictions of the models presented above, it is then relatively easy to identify the particles present in the glaze (metal or oxide), evaluate their concentration and have an idea of their size domain. Figure 8a illustrates this assertion for Cu small particles (dielectric function of the bulk and sizes ranging from 4 to 40 nm) and for copper oxide $Cu_2O$ small particles (dielectric function of bulk $Cu_2O$). These simulations clearly demonstrate that $Cu_2O$ with its weak absorption edge around 450 nm cannot achieve by itself the red color observed in our glaze. While metal Cu with its steep absorption edge around 600 nm can make it.

Figure 8b represents colors calculated with a standard illuminant D65 in the x, y, Y color space, and a 10° CIE observer, and the inserted table contains the Y values characteristic of the brightness of this color. The simulation here presented has been made for a particle concentration of 2 x$10^{-4}$, with an usual thickness of glaze equal to 150 μm. Decreasing the particle size (only taken into account in their dielectric function) makes the color evolve from red to brown and tend to desaturate (Fig. 8b). This tendency is observed at all concentrations. At increasing concentration the color becomes more saturated, but also darker, to finally become almost black above $\rho_{vol} \approx 4$ or 5 x $10^{-4}$ (this behavior is illustrated on Fig. 19a of section 6 -theoretical color chart)

### 4.1.2. Coated particles

An improvement of the Maxwell-Garnett formulation (see Bohren and Huffman [10]) allows the calculation of the absorption coefficient in the case of coated particles:

$$\alpha(\lambda) = \rho_{vol} \cdot \frac{6\pi}{\lambda} \cdot \sqrt{\varepsilon_m} \cdot \mathrm{I_m}\left[\frac{(\tilde{\varepsilon}_2-\varepsilon_m)(\tilde{\varepsilon}_1+2\tilde{\varepsilon}_2)+f(2\tilde{\varepsilon}_2+\varepsilon_m)(\tilde{\varepsilon}_1-\tilde{\varepsilon}_2)}{(\tilde{\varepsilon}_2+2\varepsilon_m)(\tilde{\varepsilon}_1+2\tilde{\varepsilon}_2)+f(2\tilde{\varepsilon}_2-2\varepsilon_m)(\tilde{\varepsilon}_1-\tilde{\varepsilon}_2)}\right] \quad (3)$$

Where $\tilde{\varepsilon}_1$ and $\tilde{\varepsilon}_2$ are now the complex dielectric functions of respectively the core and the coating and $f$ the volume ratio of the core to the total grain.

The results of the simulations presented in Fig. 9 concern the absorption coefficient of a glaze composed of Cu particles coated by a layer of $Cu_2O$ with a volume percentage of $Cu_2O$ varying from 0 (Cu grains) to 100% (oxide grains). The total grain concentration is $10^{-4}$. The thicker the oxide coating, the higher the wavelength of the absorption peak.

Such a glaze deposited on a tile with a thickness of 150 μm will produce a color (represented on Fig. 10) evolving non monotonically from red (0% oxide) to deeper red (10%), then purple (25%), blue (50%), green and eventually yellow (beyond 90%), less and less saturated and more and more luminous. In the reverse situation where the oxide is in the core and the metal in the coating (even if this case is not realistic) the effect predicted is much stronger (five times in terms of the peak wavelength displacement) and the color excursion is different: from red (0% oxide), then blue (10%), to eventually yellow for pure oxide, the tint becoming more and more luminous and desaturated. This case has not been illustrated here because it cannot produce red tiles.

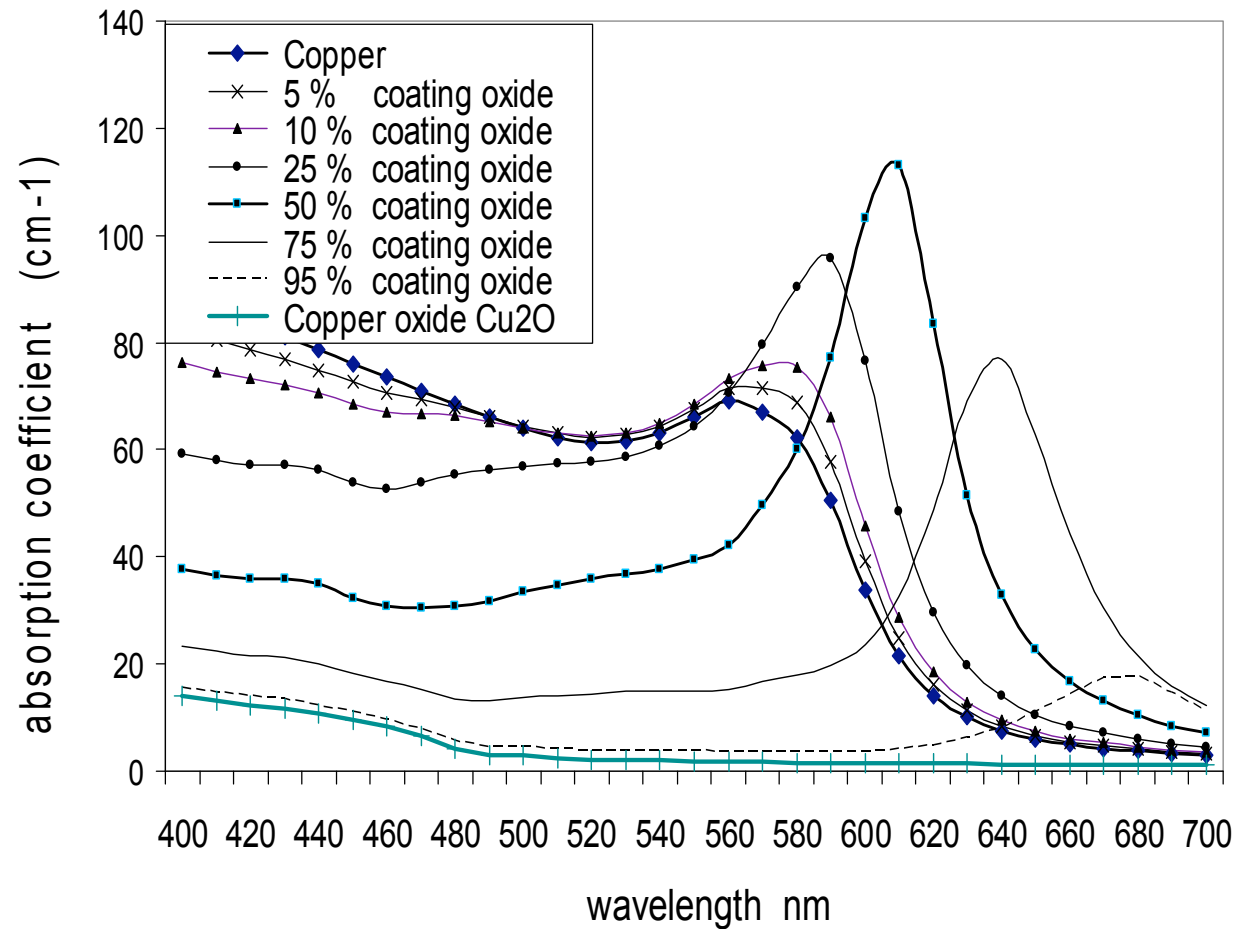

**Fig. 9** *Calculated absorption coefficient of a glaze containing oxide coated Cu particles (total $\rho_{vol}$ =$10^{-4}$) as a function of the coating volume percentage*

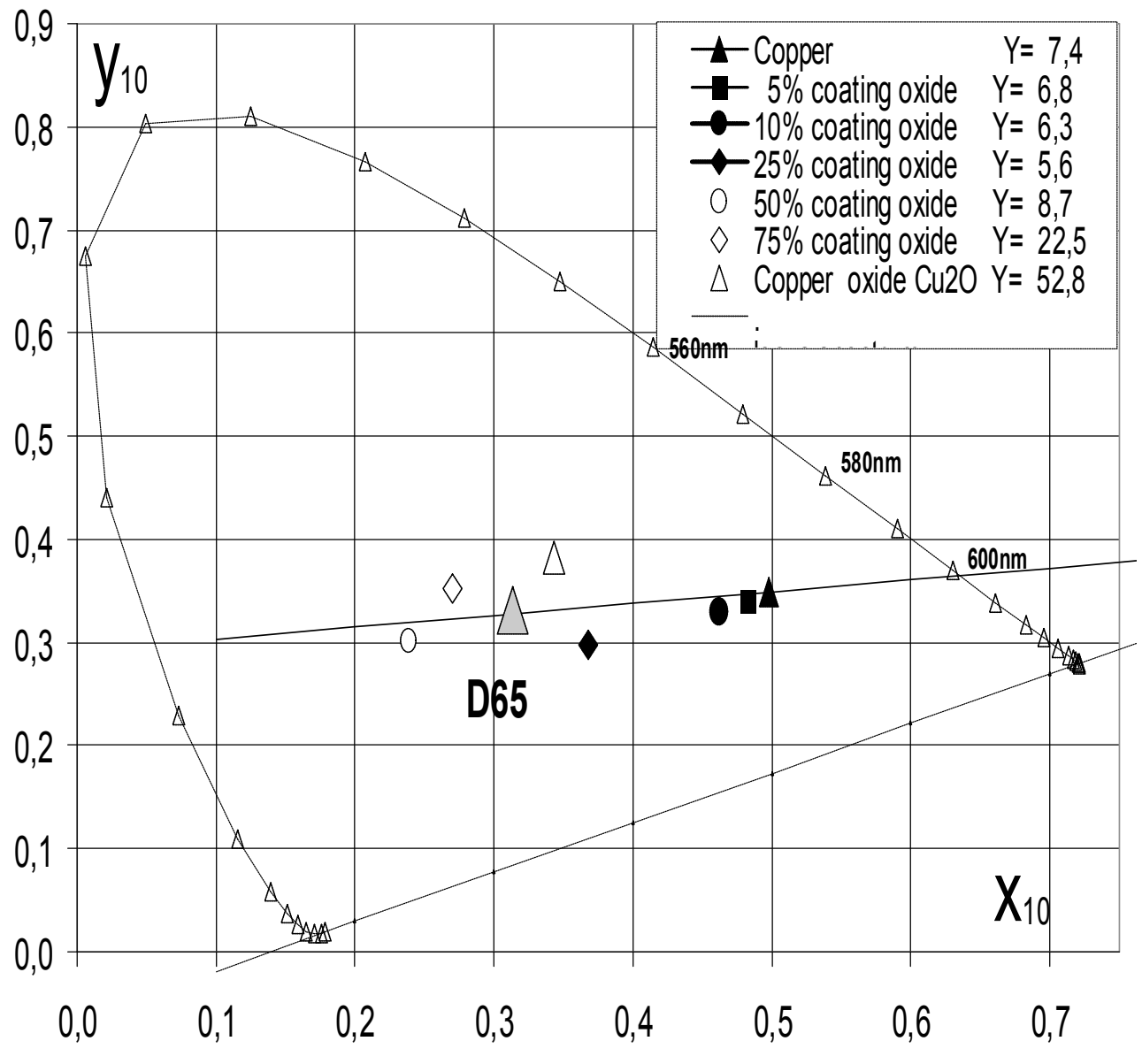

**Fig. 10** *Calculated color of a glazed tile containing oxide coated Cu particles (glaze thickness =150 µm, total $\rho_{vol}$=$10^{-4}$) as a function of the coating volume percentage*

4.1.3. Consequences on the ability of characterization :

It is first necessary to check the absence of scattering in order to validly apply the model presented above. A simple visual test consists in checking the absence of reflectance outside the specular direction of a glaze layer taken from the tile and deposited on a black velvet. The visual test can be refined by using an integrating sphere. The calculations represented in Fig. 11 show that the absorption coefficient spectra (normalized to the value of the absorption coefficient at the plasmon peak, roughly obtained at the wavelength where $\varepsilon_1 = -2\varepsilon_m$ in Eq.(2) present enough differences to allow a clear distinction between different compositions of the glaze. One can thus get:

- the nature of the particles: Cu, $Cu_2O$ or mixed
- an estimate of the particle size.

One can notice that the normalized value, following Eq 3, is independent of particle concentration.

Moreover, if the thickness of colored glaze is known (and thus the absolute value of the absorption coefficient of glaze), the suitable Eq. (2) or (3) (where alpha is proportional to $\rho_{vol}$) gives the volume ratio of particles through the relation:

$$\rho_{vol} = \frac{\alpha(\lambda.peak)measured}{\alpha(\lambda.peak)(\rho_{vol}=1)calculated}.$$

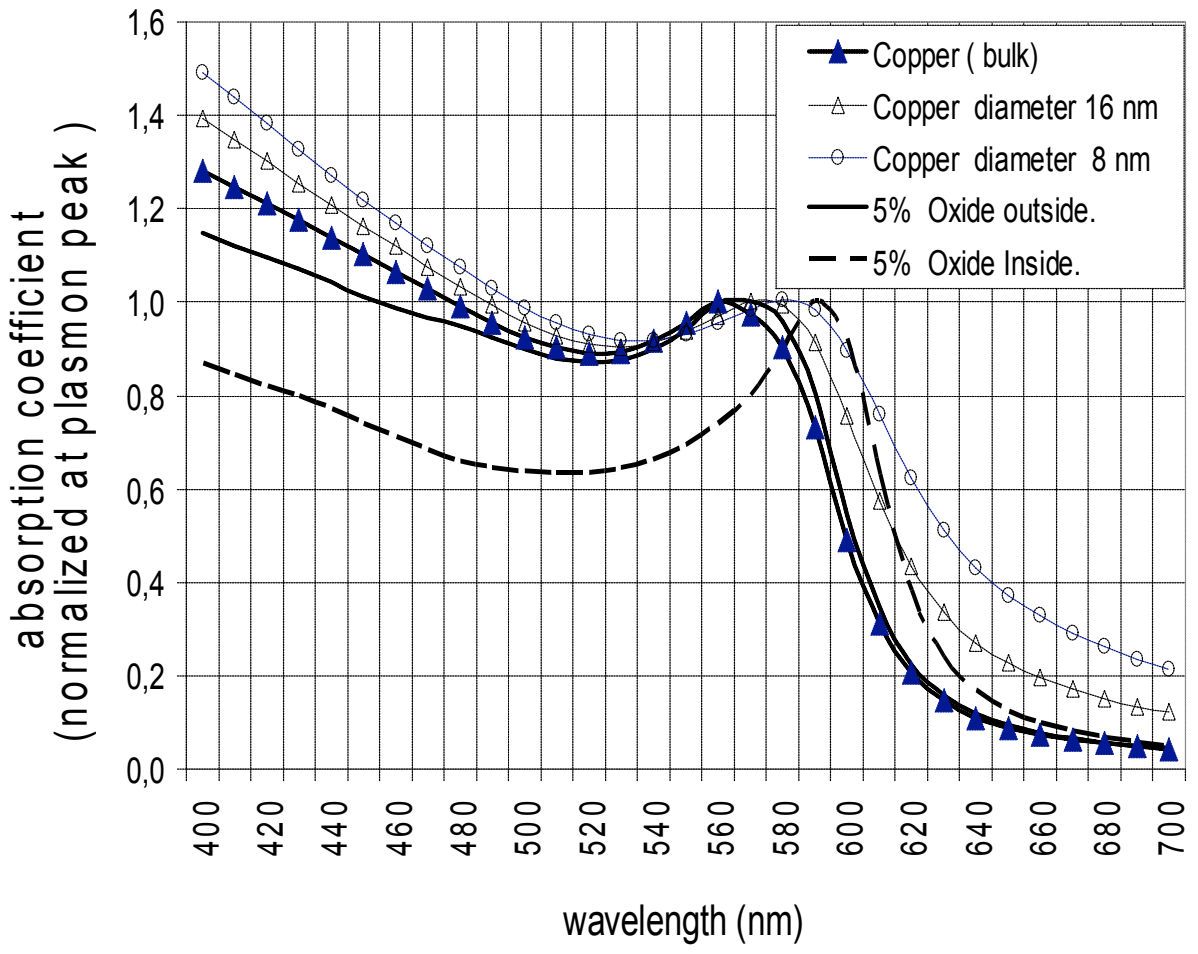

**Fig. 11** *Calculated spectral variation of the absorption coefficient of a copper glaze, normalized at plasmon peak:*
*- effect of low Cu particle size (diameter 16 nm and 8 nm)*
*- effect of 5% volume ratio of oxide (outside, or inside )*

It is clear that these characteristics, deduced from simulations, are valid provided other effects do not interfere with the parameters under consideration. It might be the case with the possible presence of other materials like Fe or Sn. Nevertheless, we estimate that their presence has no significant effect on the color of our samples, since the zones not colored in red by the copper grains are colorless (point of view of the potter). It is also confirmed by the X-ray and MET analysis (point of view of the chemist).

4.2. Large particles, light scattering

The presence of large particles induces light scattering. It makes the analysis and the interpretation much more complex. It then becomes necessary to complicate somewhat the optical model by introducing the light scattering. The concept of absorption is replaced by the concept of extinction, the sum of the effects of absorption and scattering, characterized by the corresponding coefficients:

$$\alpha_{ext}(\lambda)=\alpha_{abs}(\lambda)+\alpha_{sca}(\lambda)$$

This new coefficient acts directly on the attenuation of the specular flux. It is also involved in the spatial law of light scattering which has to be added. Mie theory can be used to calculate the absorption, $C_{abs}$, and scattering, $C_{sca}$, cross-sections of a single sphere [10] from the knowledge of its size and complex dielectric function and the refractive index of its environment. By assuming single independant scattering of an assembly of $N$ such spheres by unit volume, one can then calculate the corresponding $\alpha$ coefficients: $\alpha_{sca} = N\,C_{sca}$ and $\alpha_{abs} = N\,C_{abs}$ and deduce $\alpha_{ext}$. Some refinements of the theory have helped treat more complex cases like coated spheres and non spherical particles [10] or dependent scattering [13].

Let us first illustrate the ability of the basic Mie calculation to evaluate the size of the scatterers from the knowledge of the extinction coefficient. Fig. 12 shows the spectral variations of the extinction coefficient of a glaze composed of Cu spherical particles of diameter ranging from 80 to 170 nm, as compared to the absorption coefficient of small particles presented above (the volume fraction is $10^{-4}$). The shapes of the curves are contrasted enough to show the ability of the Mie theory to determine without too much ambiguity the particle size, assuming a single population of spheres, of low enough concentration. We recall that this parameter can be derived from a single measurement (here calculation) of the specular flux.

On the other hand, the calculation of the global optical properties of the glaze (specular and diffuse reflectance and transmittance) requires a more complex model. The Radiative Transfer approach is perfectly suited to these calculations, particularly in its 4-Flux version, that we have elected. It is thus calculated the diffuse reflectance $R_{cd}$ of a glaze layer illuminated under normal incidence

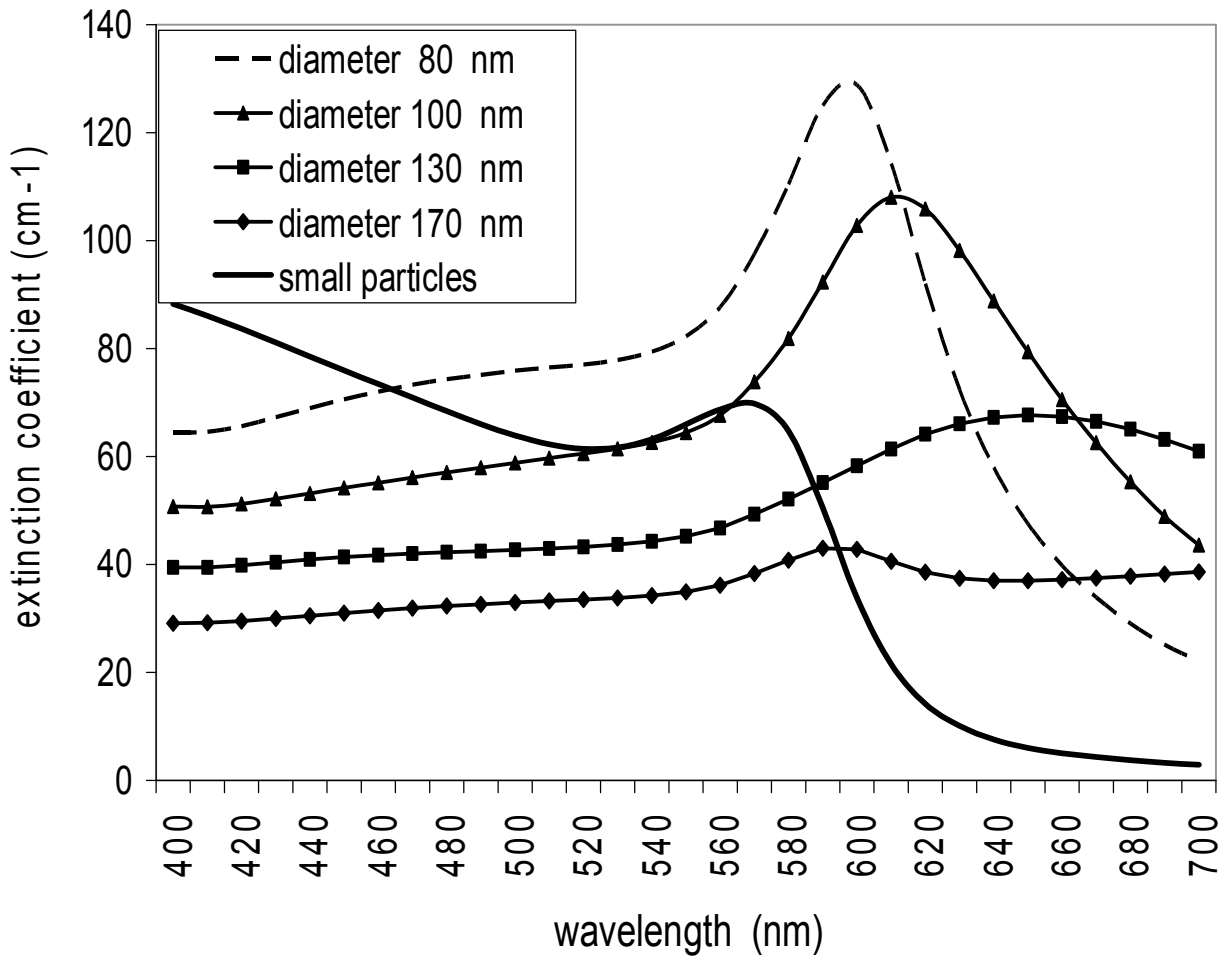

**Fig. 12** *Calculated spectral variation of the extinction coefficient $\alpha_{ext}$ (cm$^{-1}$) of a copper glaze ($\rho_{vol}$ =$10^{-4}$) as a function of the particle size*

by a collimated beam. The color observed by reflection on the layer can be directly derived from $R_{cd}$. In the case of a glaze deposited on a porcelain tile, the calculation is analogous, with different boundary conditions at the bottom of the glaze layer (the interface with the tile). We put a lambertian reflectance with a coefficient deduced from a direct measurement of the reflectance of the bare tile. In the case of a glazed tile this measurement has been corrected to take into account the difference in the refractive indices of the first medium (often called Saunderson correction [14]).

4.2.1. Glaze as a single layer

Fig. 13 shows the spectral diffuse reflectance $R_{cd}$ of a single unsupported layer, 150 µm thick, containing spherical Cu particles of diameter 100 nm, for a wide range of volume fractions from 2.5 x $10^{-5}$ to 8 x $10^{-4}$, calculated by the four-flux model. These curves show that a saturation phenomenon occurs at high Cu concentration, or more precisely (as demonstrated by other calculations not shown here) at high values of the product $\rho_{vol}$ x e (where "e" is the thickness of the glaze): let us say, for values larger than 2 x $10^{-4}$ x 150 µm). This fact proves that, if deposited on a tile, it is the glaze which -very often- entirely governs the reflectance and therefore the color. The bottom of the glaze then does not play any role.

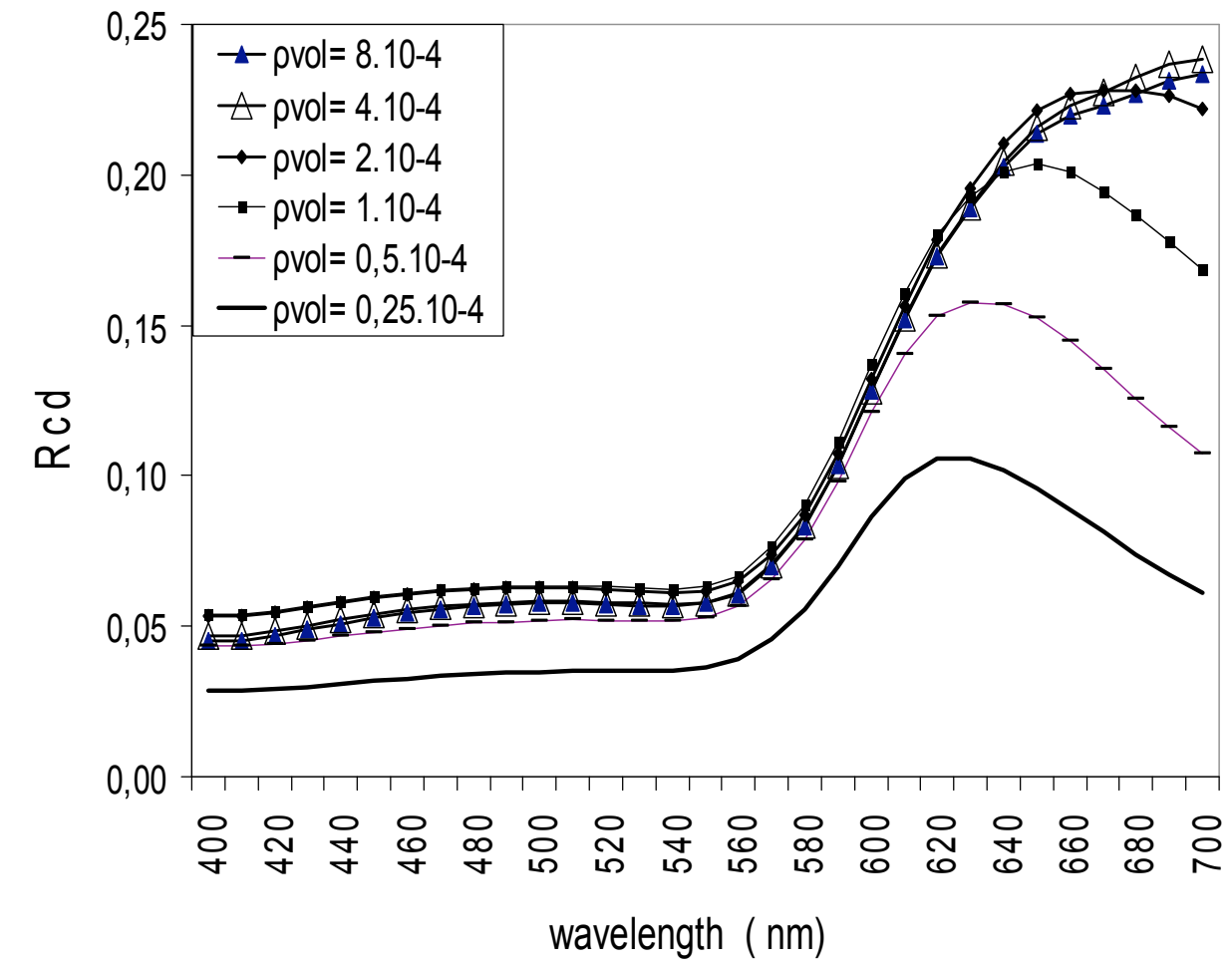

**Fig. 13** *Calculated $R_{cd}$ reflectance of a 150 µm thick layer, versus copper volume ratio $\rho_{vol}$, with a particle diameter of 100 nm*

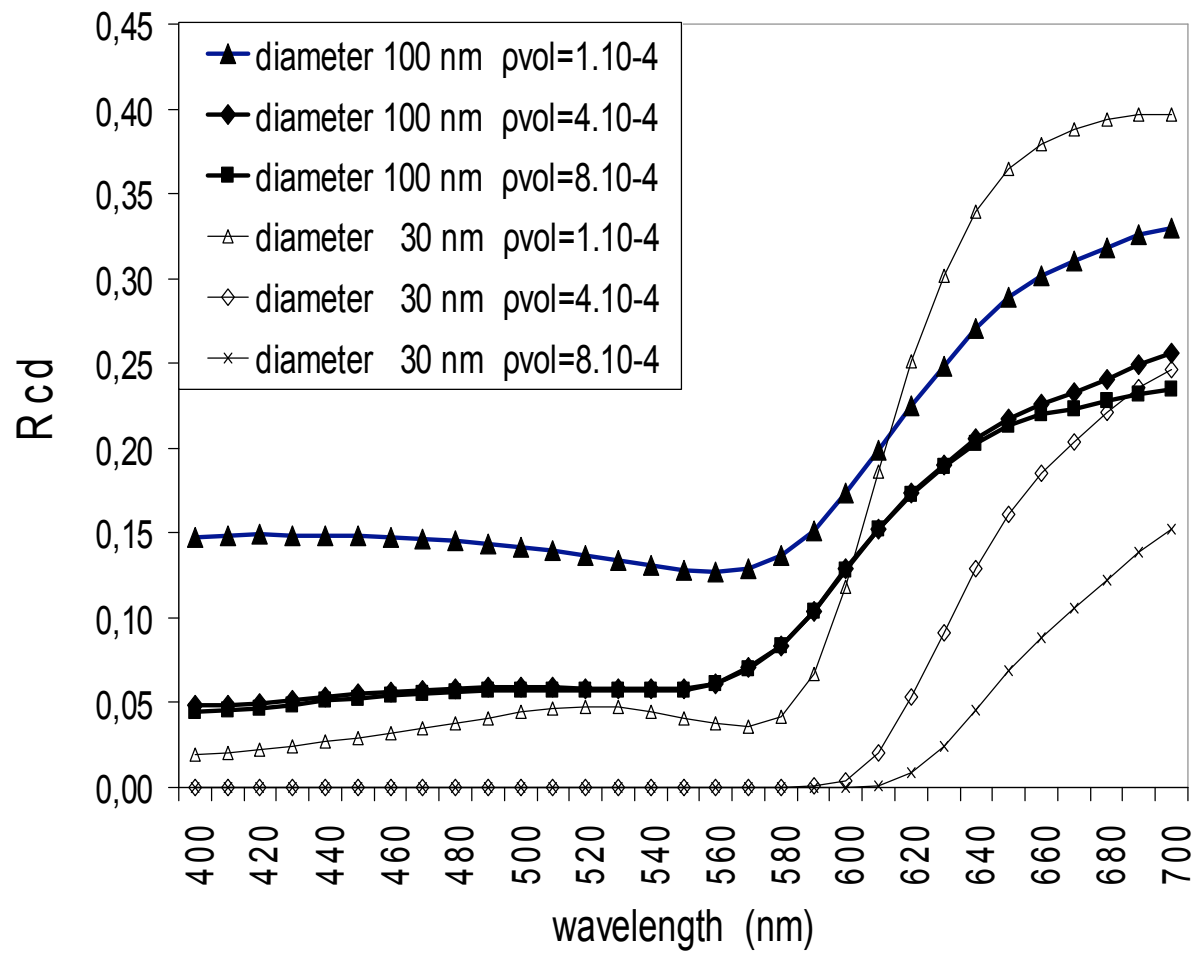

**(a)**

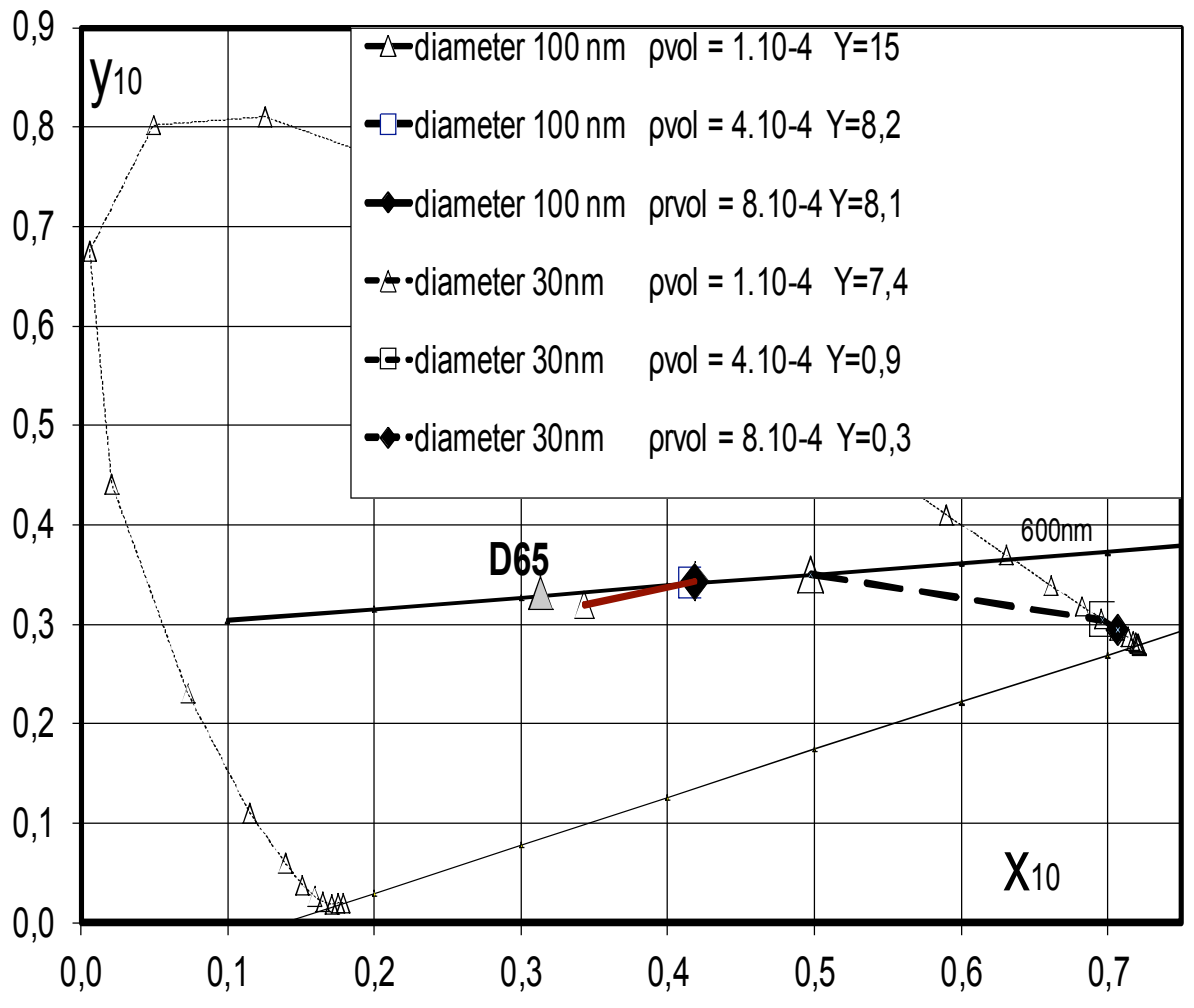

**(b)**

**Fig. 14** *Calculated $R_{cd}$ reflectance spectrum **(a)** and color **(b)** of a tile covered with a glaze 150 µm thick, for two copper particle sizes (diameter 100 nm and 30 nm) and different copper volume ratio ($10^{-4}$ to $8 x10^{-4}$)*

4.2.2. Glaze on a porcelain tile

The same calculation has been performed with the same glaze layer deposited on a porcelain tile, considering two cases associated with different particle sizes (30 nm and 100 nm), both with variable volume ratio from $10^{-4}$ to 8 x $10^{-4}$. The $R_{cd}$ curves (Fig. 14a) lead to the colors represented in Fig. 14b.

Two important results show up clearly: (i) small purely absorbing particles already produce red colors for rather low concentrations. At increasing concentration, the tint saturates and darkens. (ii) On the contrary, the glazes containing large scattering particles produce light red colors, less saturated, which tend to brown-orange at higher Cu concentration.

4.2.3. Mixture of large and small particles

By applying a mixing law to the scattering and absorption coefficient, the four-flux model can account for a medium containing a mixture of different particles (in size and/or concentration, and/or dielectric function):

$$\langle \alpha \rangle = \sum_{i} N_i C_i \;.$$

Fig. 15 shows the extinction coefficient spectra of a glaze layer, containing this time a mixture of small (diameter 30 nm) and large (diameter 100 nm) particles, as a function of the large particle ratio. In the case of a single distribution of small particles (Fig. 8, 9, 11) or large particles (Fig. 12), it was easy to evaluate both the size and the concentration of the Cu particles in the glaze layer, from the extinction coefficients deduced from specular transmission measurements on the unsupported layer. It is still true in the case of a bipopulated layer, as shown in Fig. 15, but it becomes more vague and not necessarily exploitable in all cases. Nevertheless, at a low concentration of large particles, the maximum of the curves varies little and one can make a correct evaluation of the total Cu concentration.

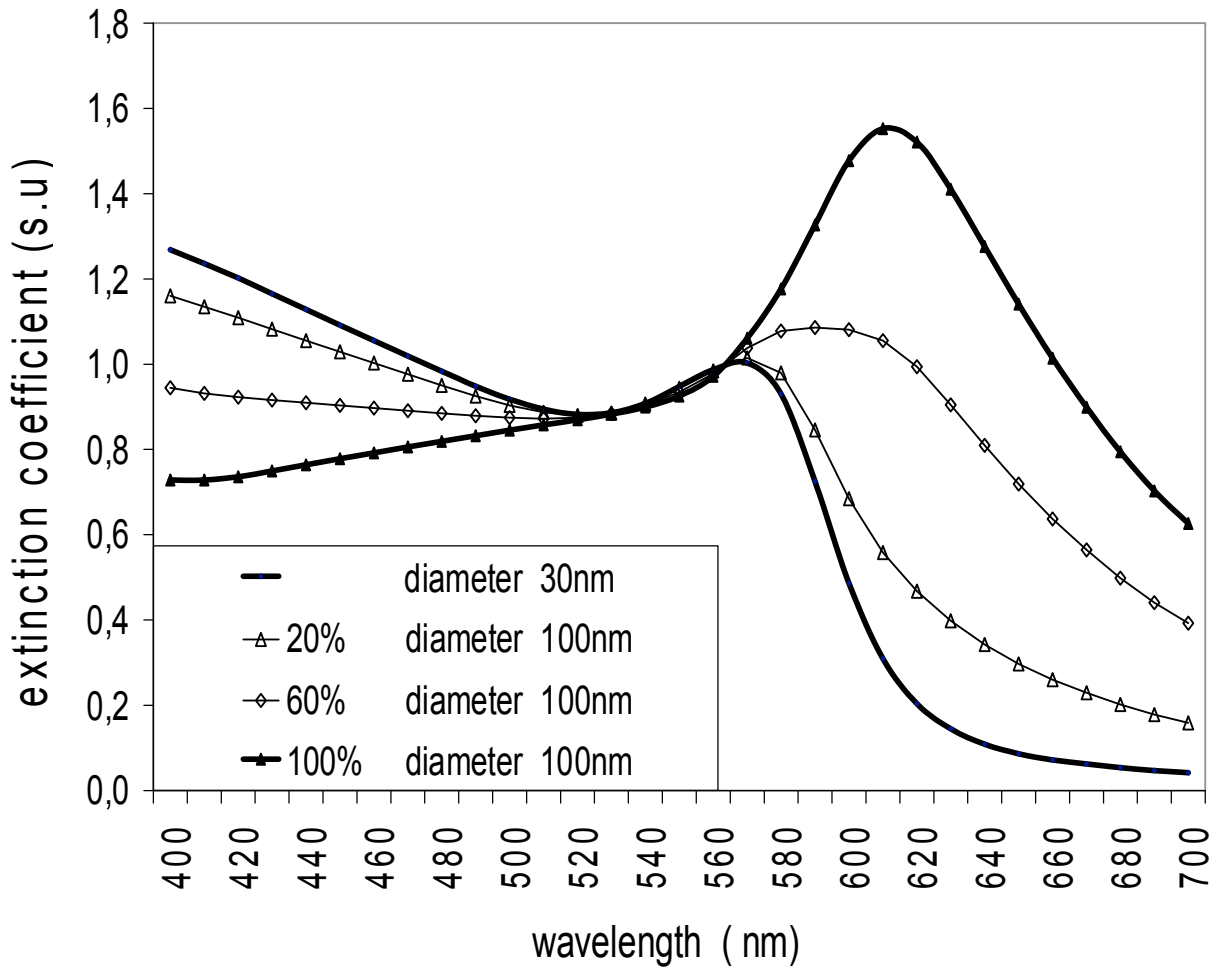

**Fig. 15** *Calculated spectra of the extinction coefficient of a bipopulated layer: small particles diameter 30 nm, large particles diameter 100 nm for different volume ratio of the large particles*

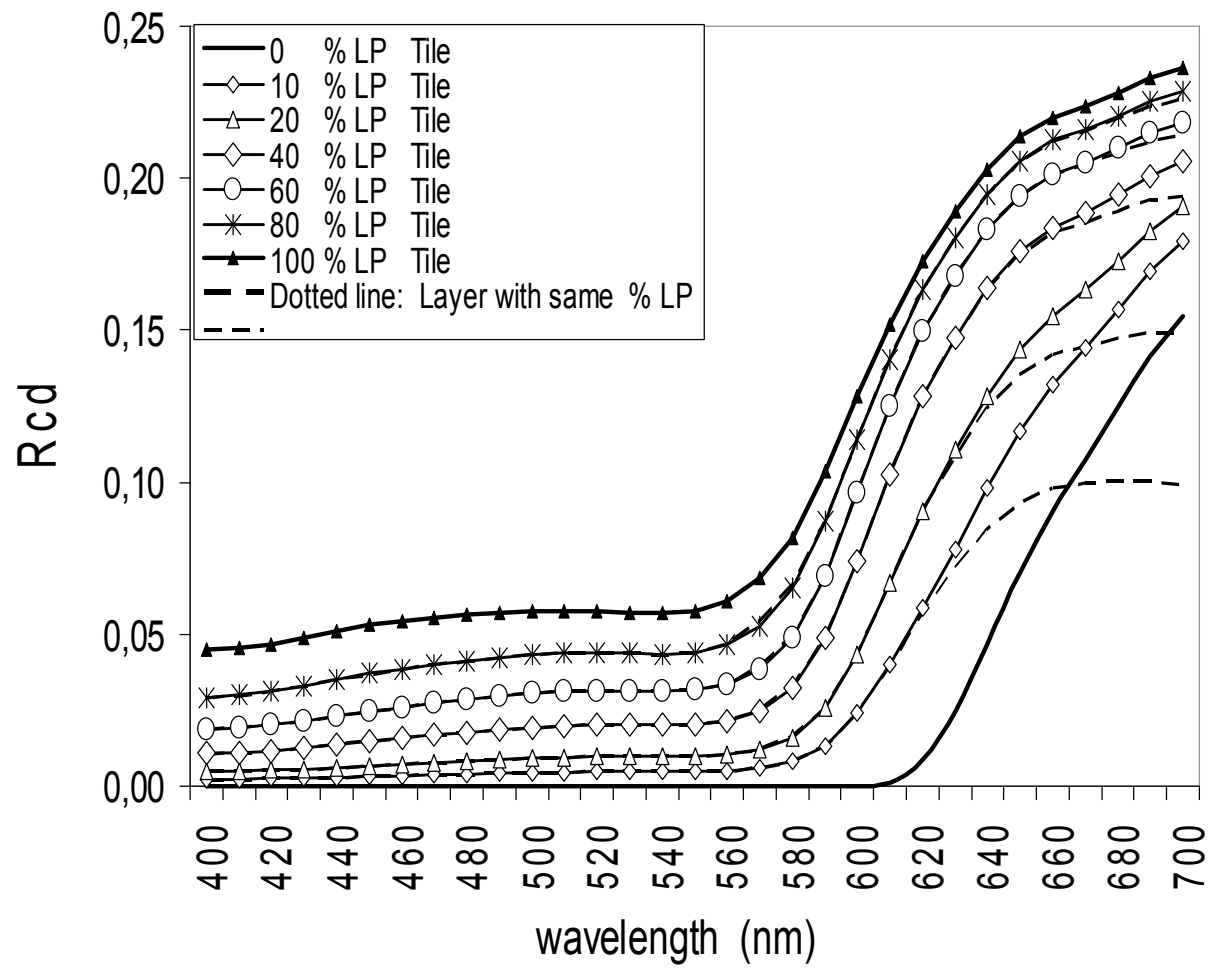

**Fig. 16** *Calculated $R_{cd}$ reflectance of a bipopulated layer (dotted line) and of a tile (full line) versus the volume ratio of large particles (LP) 100 nm diameter, thickness of the glaze 150 µm, total copper volume ratio = 8 x$10^{-4}$*

Reflectance measurements on tiles covered with these glazes (as simulated in Fig. 16) are still more difficult to use precisely, because they depend on the absorption and scattering coefficients, as well as on the Cu concentration, in a non linear way. They are however essential to identify the presence of scattering and useful to specify the concentration and therefore the role of the scattering particles. Qualitatively, the general trends observed in Fig. 14 on tiles with glazes composed of single size particles, are confirmed in the case of a mixture. Abundance of small Cu particles leads to a decrease in reflectance below 500-600 nm and shifts towards the red the high reflecting zone, producing therefore a red color more saturated. While large particles desaturate the color and increase the brightness.

## 5- Final interpretation

The optical arguments developed under a theoretical (or modeling) point of view in section 3, supported by the structural characterizations from the MET, will govern the interpretation of the optical measurements made on the tiles and on the glaze unsupported layers.

### 5.1. Transmittance measurements

Following the principle of Eq. 1, the collimated transmittance $T_{cc}$ of the glaze layers allows the calculation of the extinction coefficient. The spectral curves in Fig. 17 are normalized to the value at their peak around 570 nm. By comparison with the curves of Fig. 15 related to bi-populated media, the curves of Fig. 17 clearly show, assuming an average diameter of the scattering particles equal to 100 nm, that their volume ratio is close to 20%, and in any case much less than 60%, due to the spectral position of the peak. One can therefore get an evaluation of the total volume fraction of Cu in the glaze, by comparing the absolute values of $\alpha_{ext}$ in the glaze and in bulk Cu serving as a reference.

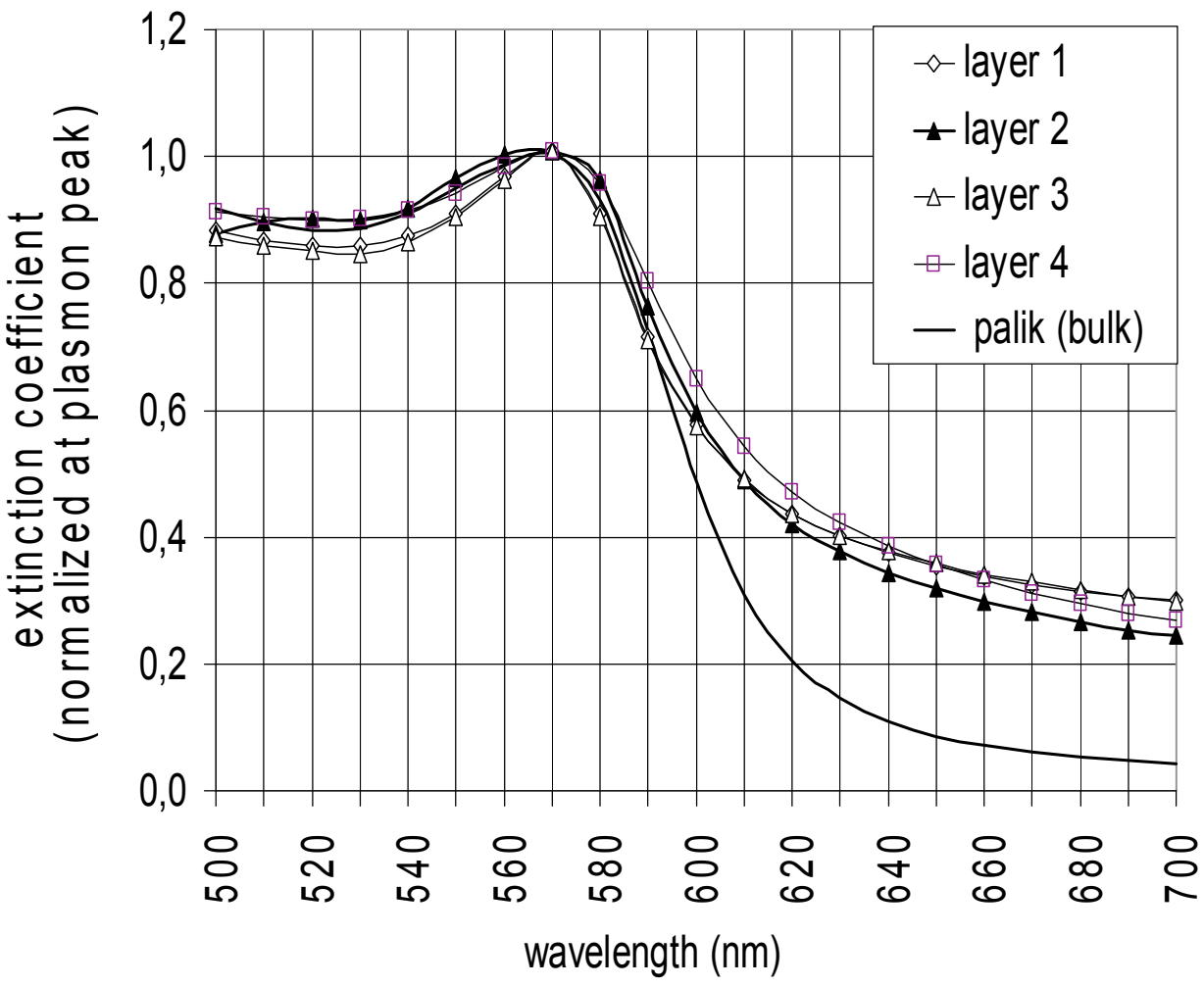

**Fig. 17** *Extinction coefficient of B23 layers, normalized at plasmon peak, deduced from Tcc measurements (see Fig. 3a), compared to calculated one (bulk)*

One gets values close to 9 x $10^{-4}$ for the three first layers and 12 x $10^{-4}$ for the deepest one. Let us remind that a value of 9 x $10^{-4}$ corresponds to a total conversion of the Cu carbonate (put in the batch of the glaze with a massic ratio of 6 x $10^{-3}$) into metallic Cu. These results show that the glaze looks fairly homogeneous in its depth, with a slightly higher concentration in the layer closer to the tile body. This observation is coherent with the reflectance measurements of Fig. 3b.

5.2. Reflectance measurements

The measurements of the diffuse reflectance $R_{cd}$ (Fig. 18) on the tile and on the glaze layer number 2 are coherent with the predictions of the four-flux model, by assuming that the values of the main parameters are those deduced from the previous results: total Cu volume fraction equal to 8 x $10^{-4}$, glaze thickness 150 µm, small particles non absorbing, large particle diameter 100 nm, and confirms and makes more precise the ratio of large particles: 0.20-0.30. It is worth noticing that this optical model allows thus an evaluation of this ratio which is not possible with the TEM.

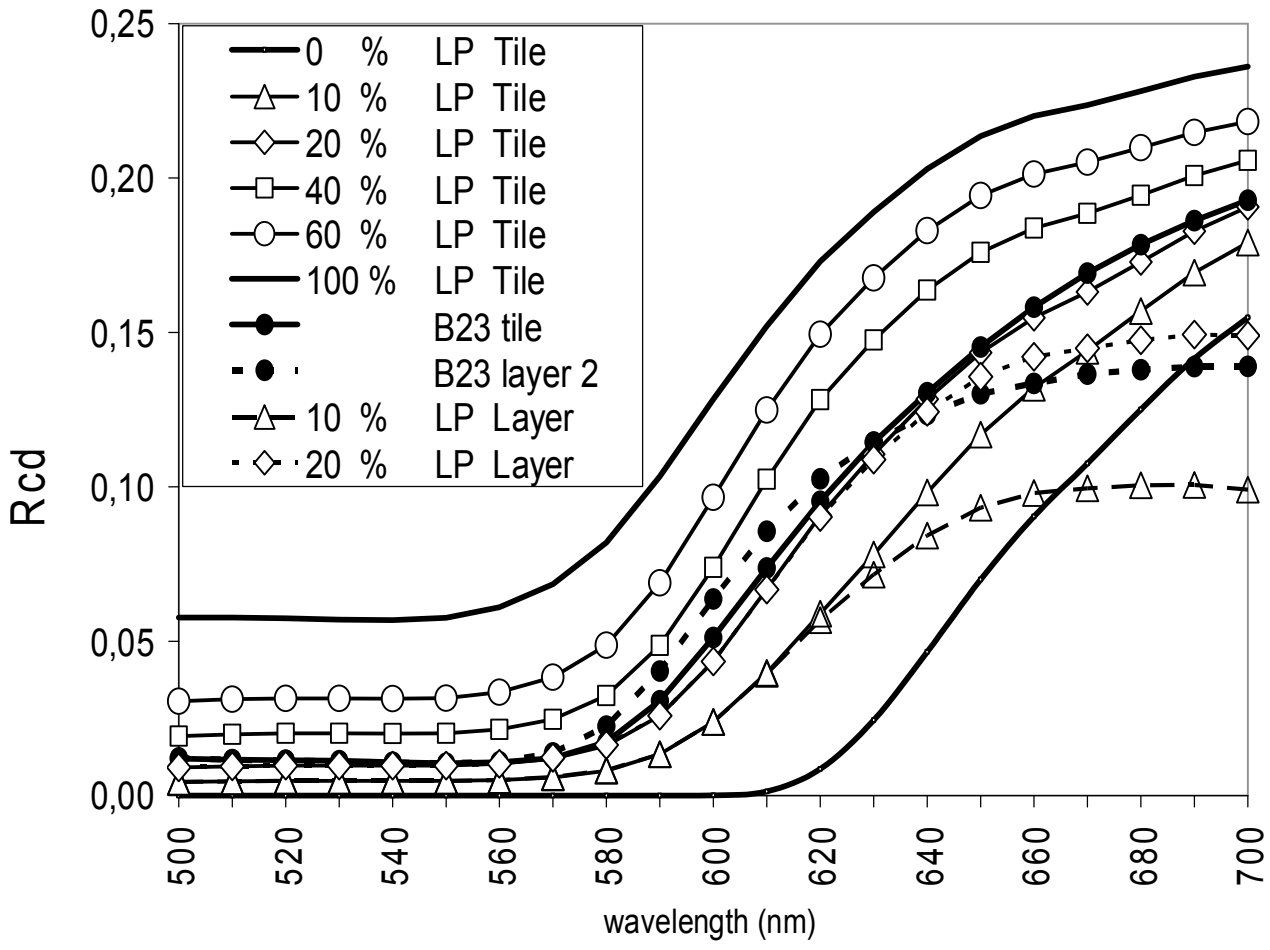

**Fig. 18** *Tile B23: Comparison of the measured reflectance $R_{cd}$ (see Fig. 3b) with the one calculated with a four-flux model (see Fig.16: large particles diameter 100 nm, thickness of the glaze 150 µm, total copper volume ratio = 8 x$10^{-4}$):*
- *Black dots = measured values, white marks = calculated values*
- *Full lines: whole tile*
- *Dotted lines: layers (measured values of B23 layer 2, and calculated values for 10% and 20% of large particles ratio)*

5.3. First conclusions

At this step of the modeling, one can already draw several conclusions:

a) - The whole set of experimental measurements on both the thin unsupported layers and the tiles is coherent with the modeling of a glaze containing two families of nanoparticles, on the basis of the radiative transfer approach (four-flux model).

b) - Measurements and simulations allow thus a correct description of the morphology of the colored layer: (i) no particles of size smaller than 10 nm; (ii) relatively homogeneous dispersion of the particles in the depth of the layer with a slightly higher density at the bottom of the glaze; (iii) almost total conversion of the Cu carbonate into Cu during the firing; (iv)volume ratio of large particles equal to 0.20-0.30.

c) - The reflectance and the color of the tile clearly result from the combined effects of the presence of these two families.

The low value of the volume fraction of the metallic particles (less than $10^{-3}$) has to be pointed out again. If they were on a cubic network, their distance would be 1.5 µm for the large particles and 0.25 µm for the small ones. This remark justifies the relative efficiency of the optical spectroscopy over the TEM for evaluating this parameter

## 6. Theoretical color chart

The optical study of a substantial number of tiles representative of the broad Cu red palette has given a set of reflectance and transmittance curves (specular and diffuse) and volume fraction values coherent with the presence of the two families (large and small Cu particles) described above. Assuming that our optical model has been validated by this study, it is consistent to try to use it for predicting a theoretical color chart of the reachable colors. We will establish this chart in a first step with large particles of diameter 100 nm. We will then study the effect of increasing their size.

6.1. Large particles 100 nm

6.1.1. Brightness Y

The brightness Y of a tile covered with a glaze, 150 µm thick, with a Cu volume fraction 8 x $10^{-4}$ (usual value in the practice of a potter), is roughly proportional to the volume fraction of the large particles and quasi-independent of the Cu concentration as soon as it is larger than 4-5 x $10^{-4}$ (Fig. 19a). The same calculation shows that the tile looks almost black (Y value around 1) below a ratio of large particles of 5-10%. Indeed, in this case, the main part is played by the small absorbing particles

6.1.2. Chromatic components x, y

Fig. 19b shows that, similarly to the brightness, the chromatic components x and y strongly depend on the large particle ratio. The behavior already mentioned above is again observed here. At increasing volume ratio: desaturation of the hue, higher brightness and hue angle departing from red except for low copper ratio (see curve 2 x $10^{-4}$). x and y become also independent of the Cu concentration beyond 4-5 x $10^{-4}$.
However, at low concentration, this parameter becomes crucial and governs both (i) the scattering of the glaze (as illustrated by Fig. 13 in the case of large particles alone) and (ii) the flux reflected by the porcelain body at the bottom of the glaze, which becomes large when the absorption of the glaze weakens. Fig. 19 also shows that in the extreme case of a glaze only composed of small particles it is their concentration which governs the color (dashed curve on Fig. 19b). It becomes darker and redder whith a higher concentration, and tends to black (Fig. 19a) when the copper volume fraction reaches 4 to 5 x $10^{-4}$.

6.2. Effect of the size of the scattering particles

In order to illustrate this point, we have chosen a high Cu concentration (the thickness of the glaze is still 150 µm) (Fig. 20). Within this domain, usual in the practice of the potter, we have already mentioned that the color is weakly sensitive to the Cu concentration.

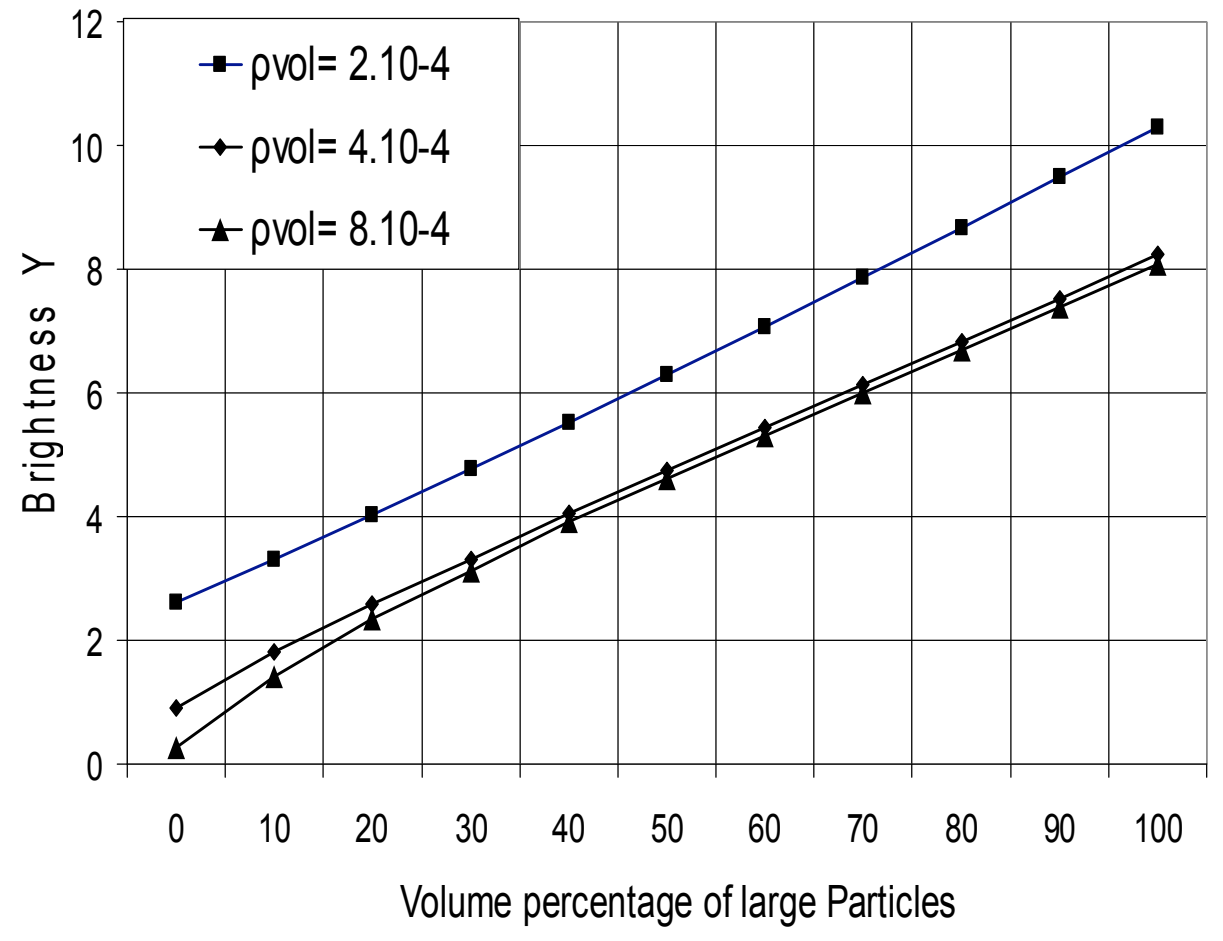

(a)

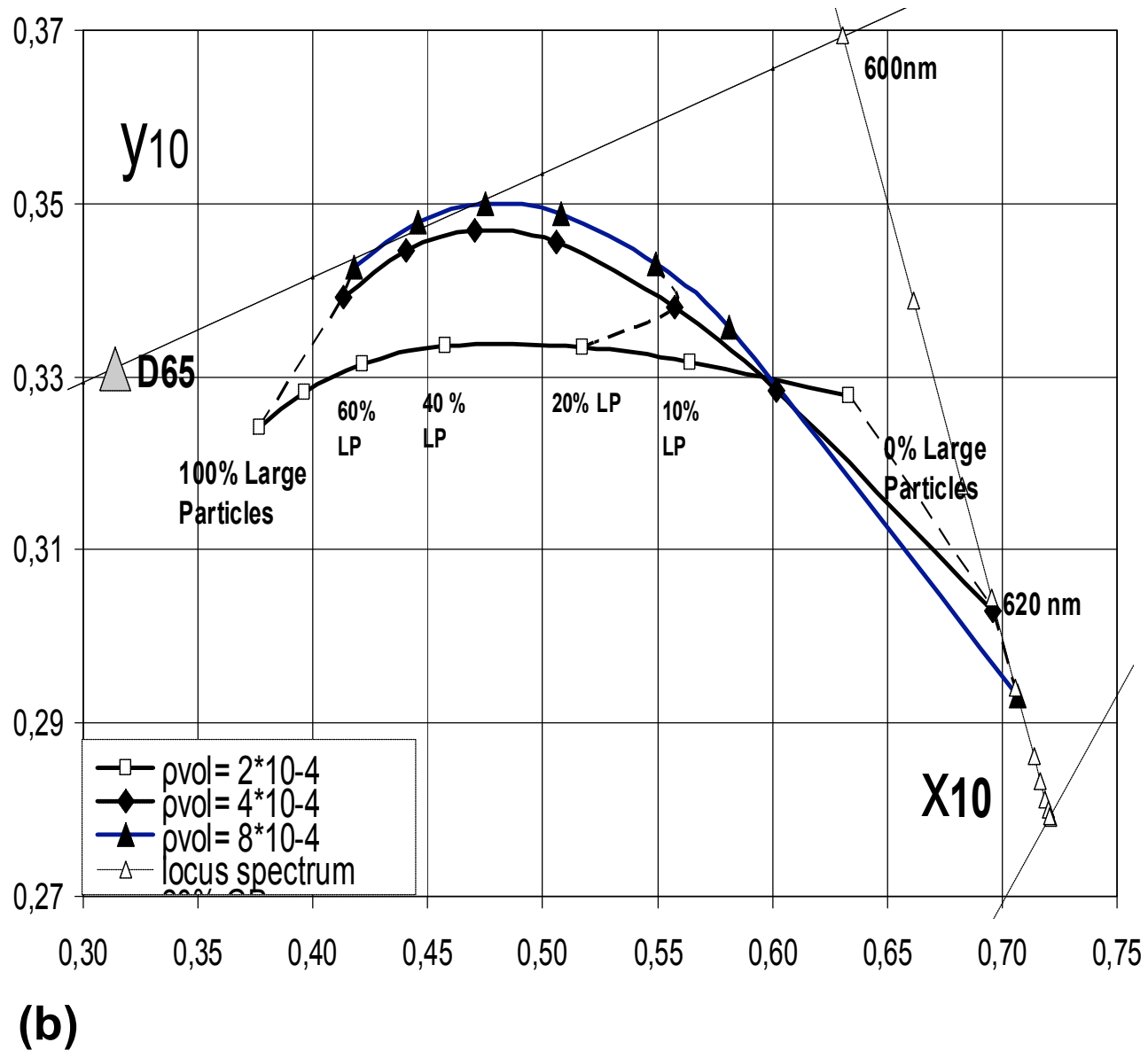

(b)

**Fig. 19** *Calculated variations of* ***(a)*** *brightness Y and* ***(b)*** *color coordinates x, y, of a tile, versus the volume ratio of large particles, for different copper concentrations (glaze thickness 150 µm)*

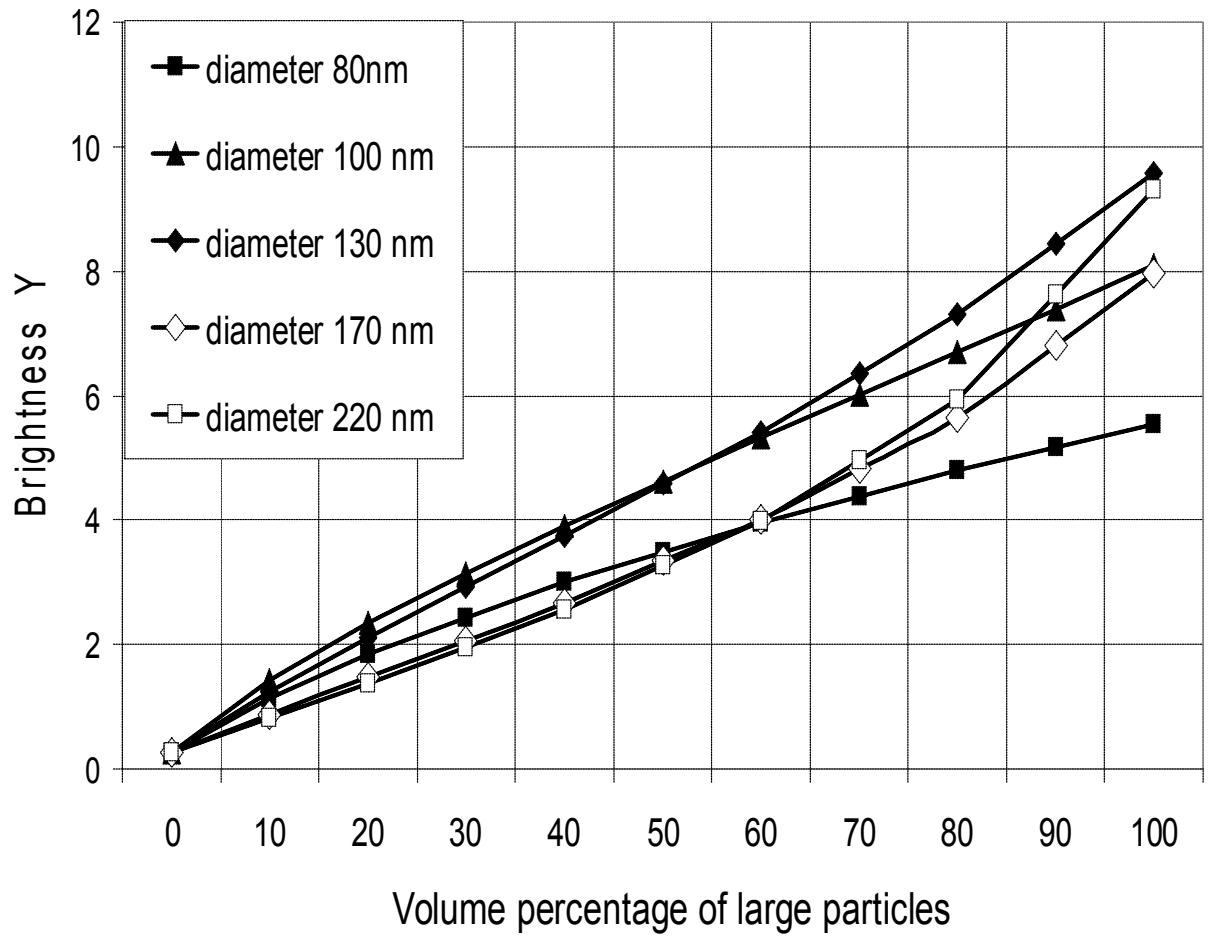

(a)

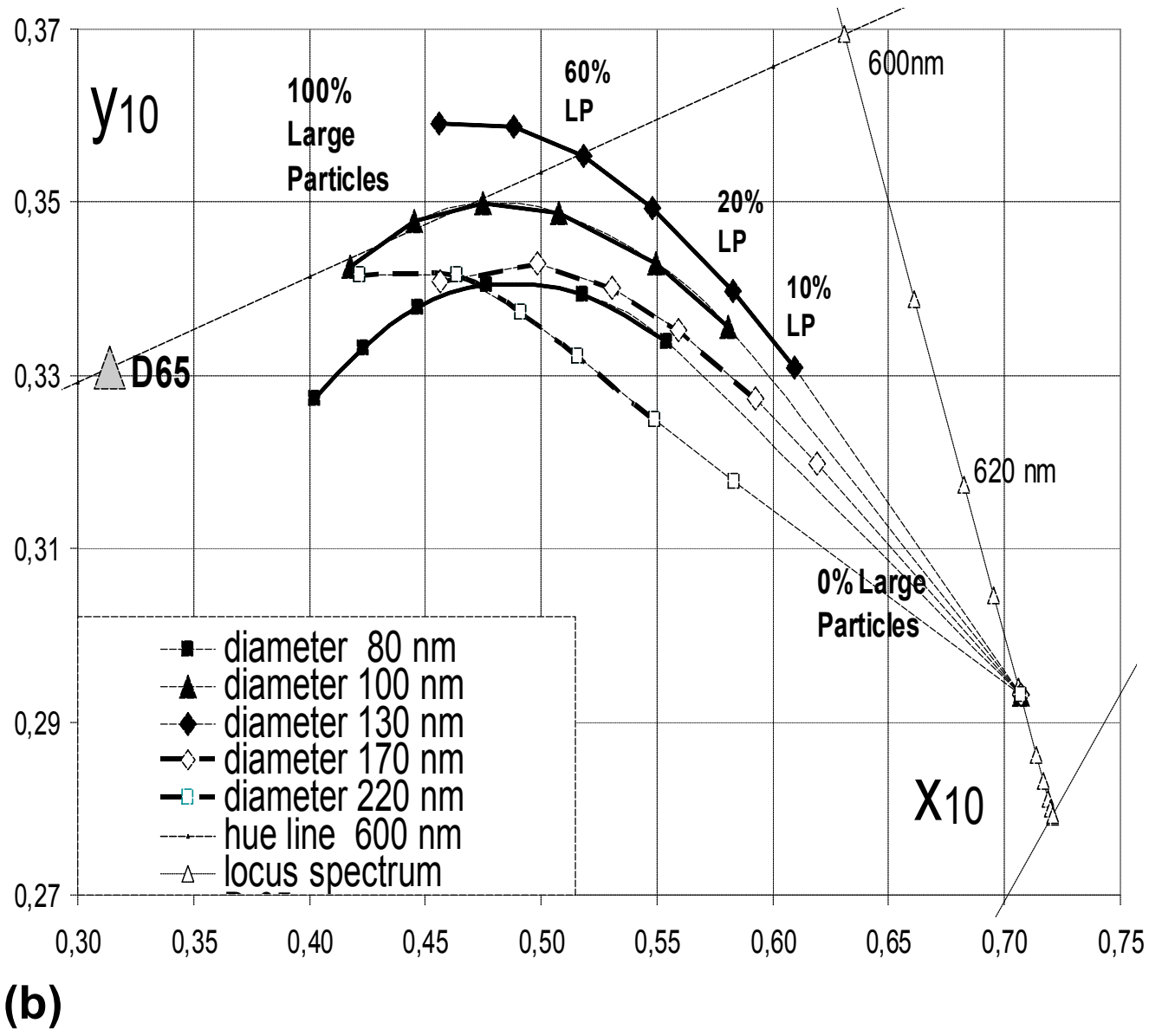

(b)

**Fig. 20** *Calculated effect of the size variation of large particles on* ***(a)*** *brightness Y and* ***(b)*** *color coordinates x, y of a tile, with variation of large particles volume ratio (copper volume ratio = 8 x$10^{-4}$, glaze thickness 150 µm)*

One can also notice that for a ratio less than 10-15%, depending on the size, the tiles become almost black (Fig. 20a).

6.2.1. Influence of the large particles ratio

At any size of the scattering particles, their relative volume ratio still plays a major part. The brightness, saturation and hue roughly vary as already observed in Fig. 19 (as a function of this parameter).

6.2.2. Influence of increasing the size of the large particles

Nevertheless, all else being equal, increasing the size of the particles does not always act in the same way on the color (Fig. 20b). At the beginning (80 to 130 nm) it increases the brightness and weakens the saturation and the color becomes less red. The behavior is then reversed up to more than 200 nm.

**7. Discussion and prospects**

7.1. We have here highlighted the crucial part of the presence of two families of particles in producing the color. It implies the existence of two distinct phenomena and can help further research for a better control of color variations in glazes. It is also consistent with the fact that a glaze is basically non homogeneous (at different scales) due to the firing process on a tile or a pot, which prevents a good mixing between different melting zones, offering thus possibilities of different germination and growth processes. In contrast, we must recall that copper red glasses - for which homogeneity is a major goal - are composed of only one family of small absorbing metal particles .

7.2. All the simulations and the measurements presented in this paper concern samples considered as homogeneous (especially in the plane of the layer) at the scale of the light beam. However, actual glazes present a lot of macroscopic heterogeneity: (i) colorless zones dispersed in the red; (ii) darker veins or heterogeneities. All these "defects" make the beauty of the ceramics. Their optical effect could be separately easily modeled. But their linear combination will never account for that beauty.

7.3. We have not found in the literature, nor in tests , red color in copper glaze due to oxide coated Cu particles. Moreover, this study tends to prove its impossibility with a significant coating ratio.

7.4. Blue color can sometimes be obtained during the elaboration of copper reds, apparently due to surface layer effects. According to the intensity of the effect, one can obtain different hues: greyish, violine, purple and pale or strong blue. This effect deserves by itself a special study. In the same way, some authors [8] have emphasized the importance of a thin red surface layer for the final quality of color for the pieces they have studied. The influence of this kind of layer could be evaluated or modelised with the help of the results or methods of the present study.

**8. Conclusion**

- Copper red glaze pottery have been elaborated by a ceramist. A careful analysis of the structure and composition of these glazes, by MET and EELS followed by an optical characterization and an optical modeling using the radiative transfer approach (four-flux model), have enabled the first full explanation of the origin of this optical effect: the presence of two families of Cu particles in the vitreous matrix. The first, purely absorbing, of diameter 10 to 50 nm essentially creates color by a substractive process. The second, due to its larger diameter, 100 nm or more, mainly acts on color by scattering of the visible light. Both act competitively in the layer.
- The presence of copper oxide has not been detected in the glazes. Moreover, we did not find any evidence that Cu particles with a $Cu_2O$ coating may play a role in red color of copper glazes. Moreover, simulations tend to prove that $Cu_2O$ grains cannot produce red color .
- The large particles, even at low concentration, play a major role in achieving the color, due to their scattering effect. Their absence (or a too low concentration) leads to almost black pottery. Their presence enhances the brightness of the color and tends to desaturate the hue. Beyond a given concentration and size, their scattering mean free path becomes smaller than the thickness of the glaze layer, making the color independent of the glaze thickness. Their volume relative proportion (over all the Cu particles) is a major parameter of the color.

- In conclusion, two distinct phenomena of germination and growth of Cu particles occur in the glaze during the firing, which eventually govern the color of the pottery. They still remain an important point of investigation.

- The color chart that we have calculated gives an idea of the total palette which might be theoretically

covered by the glazes colored by Cu nanoparticles. This chart also makes clearer the role played by the large particles.

**Aknowledgements:** P.A.C. wishes to thank Daniel Steen (HEI, Lille) and Wahib Saikhaly (CP2M, Marseille) for their help respectively in tile optical measurements, and in EELS studies